\documentclass[utf8,bibauthoryear,printer]{aa}
\usepackage{graphicx}
\usepackage{txfonts}
\usepackage{siunitx}
\usepackage[utf8]{inputenc}
\usepackage{natbib}
\usepackage{caption}
\usepackage{subcaption}
\usepackage{amsmath}
\usepackage{array}
\usepackage{bm}
\usepackage[pdftex, hidelinks]{hyperref}
\begin{document}

   \title{Radiative Rayleigh-Taylor Instability and the structure of clouds in
     planetary atmospheres} 

   \author{P. Tremblin  \inst{1}\thanks{\email{pascal.tremblin@cea.fr}}
     \and H. Bloch      \inst{1}
     \and M. Gonz\'alez \inst{2}
     \and E. Audit      \inst{1}
     \and S. Fromang    \inst{2}
     \and T. Padioleau  \inst{1}
     \and P. Kestener   \inst{1}
     \and S. Kokh       \inst{3}
   }

   \institute{Maison de la Simulation, CEA, CNRS, Univ. Paris-Sud, UVSQ,
     Universit\'e Paris-Saclay, F-91191 Gif-sur-Yvette, France\
     \and  AIM, CEA, CNRS, Universit\'e Paris-Saclay, Universit\'e de Paris,
     F-91191 Gif-sur-Yvette, France\
     \and DEN/DANS/DM2S/STMF, CEA Saclay, 91191 Gif-sur-Yvette, France
   }

   \date{Received \#\#\# \#\#\, 2019; accepted \#\#\# \#\#, 2019}

  \abstract
   {}
   {Clouds are expected to form in a wide range of conditions in the atmosphere
     of exoplanets given the large range of possible condensible
     species. However this diversity might lead to very different small-scale
     dynamics depending on radiative transfer in various thermal conditions: we
     aim at providing some insights into these dynamical regimes.}
   {We perform an analytical linear stability analysis of a
     compositional discontinuity with a heating source term that depends on
     composition. We also perform idealized
     two-dimensional (2D) simulations of an opacity discontinuity 
     in a stratified medium with the code \texttt{ARK}. We use a two-stream grey
   model for radiative transfer and explore the brown-dwarf and earth-like
   regimes.}
   {We reveal the existence of a Radiative Rayleigh-Taylor Instability (RRTI
     hereafter, a particular case of diabatic Rayleigh-Taylor instability)
     when an opacity discontinuity is present in a stratified medium. This
     instability is similar in nature to diabatic convection and relies only on
     buoyancy with radiative transfer heating and cooling. When the temperature
     is decreasing with height in the atmosphere, a lower-opacity medium on
     top of a higher-opacity medium is dynamically unstable while a higher-opacity
     medium on top of a lower-opacity medium is stable. This stability/instability
     behavior is reversed if the temperature is increasing with height.}
   {The existence of the RRTI could have important implications for the
stability of the cloud cover of a wide range of planetary atmospheres.
In our solar system, it could help explain the formation of mammatus
cloud in Earth atmospheres and the existence of Venus cloud deck.
Likewise, it suggests that stable and large scale cloud covers could be
ubiquitous in strongly irradiated exoplanets but might be more patchy in
low-irradiated or isolated objects like brown dwarfs and directly imaged
exoplanets.}

   \keywords{(Stars:) brown dwarfs, Earth, planets and satellites: atmospheres,
     gaseous planets, terrestrial planets}

   \titlerunning{Opacity-Jump Instability}
   \authorrunning{P. Tremblin}

   \maketitle
%

\section{Introduction} \label{sec:introduction}

With more and more observations of exoplanet atmospheres with
transmission spectroscopy \citep[e.g.][]{sing:2016}, a growing concern about clouds
has been spreading in the exoplanet community. Clouds may lead to ``flat''
spectra of the atmosphere of super-Earths/mini-neptunes \citep{kreidberg:2014}
potentially reducing our ability to get information about the gas phase but also
providing clues of where and how condensible materials are forming and evolving.

Cloud modeling remains a very important challenge in exo-planetology given the
complexity of the phenomena that link chemistry with complex microphysics,
radiative transfer, and hydrodynamics. Several complementary approaches have
been developped in the past few years regarding this challenge: global 1D models
with either simplified \citep[e.g.][]{fortney:2008,tan:2019} or
complex \citep[e.g.][]{helling:2019} microphysics; and 3D global circulation
models (GCMs) with passive \citep[e.g.][]{parmentier:2016} or active
\citep[e.g.][]{lines:2018} clouds. Early studies such as \citet{gierasch:1973}
and more recent ones \citep[e.g.][]{tan:2021a,tan:2021b} have also identified the key
role of radiative instabilities at a global scale, in planetary atmospheres
(Venus, Jupiter), and in the atmosphere of exoplanets and brown dwarfs.

Yet all these approaches use a
simplified approach to hydrodynamics, either because they are 1D, or because
they are global and cannot capture properly small-scale, non-hydrostatic buoyancy effects. We therefore
propose in this paper another complementary approach based on theory with the
inclusion of source terms in the Rayleigh-Taylor stability analysis similarly to the
recent theoretical development of diabatic convection proposed in
\citet{tremblin:2019}. The classical (incompressible) Rayleigh-Taylor
  instability is triggered by a jump in density with a heavy fluid on top of
  a light fluid \citep[see][]{chandrasekhar:1961}, but cannot account for the
  stabilizing/destabilizing effect of source terms similarly to Schwarschild convection.
  We then use local small-scale
simulations in order to properly study the interplay between buoyancy and
radiative transfer with opacity dicontinuities and its impact on the structure
of clouds.

\citet{tremblin:2019} recently proposed a new paradigm for the development of
convective motions in the presence of compositional and thermal source terms,
i.e. diabatic convection. This paradigm can descibe many convective systems,
such as moist convection in Earth atmosphere, thermohaline convection in Earth
oceans, and CO/CH$_4$ radiative convection in the  atmosphere of brown dwarfs
and giant exoplanets. In this paper we study in a similar way the
dynamical behavior of an opacity discontinuity between a ``higher-opacity'' and
``lower-opacity'' medium subject to radiative heating and cooling. Such a
discontinuity could be unstable/stable due to the impact of radiative transfer on
buoyancy similarly to the impact of thermal source terms in diabatic
convection. This Radiative Rayleigh-Taylor Instability (RRTI hereafter) is
however an interface
instability different from diabatic convection similarly to the difference
between the standard Rayleigh-Taylor instability and Schwarschild convection in
the adiabatic case.


In Sect.~\ref{sec:rrti} and Appendix~\ref{app:rrti} we present the linear
stability analysis predicting the RRTI growh rate and we present our numerical
setup in Sect. \ref{sec:model}. In Sect.~\ref{sec:bd-model}, we propose a first
idealized study in the brown-dwarf regime and provide a simplified
understanding of the RRTI mechanism. In Sect.~\ref{sec:earth-model}, we study the
impact of the temperature gradient of the atmosphere on this instability for an
Earth-like idealized setup. In Sect.~\ref{sec:clouds}, we discuss the possible
implications of this instability on the cloud structure of different
atmospheres. We then reach our 
conclusions in Sect.~\ref{sec:conclusions} and discuss the possible expectations
for future exoplanet observations.

\section{Linear stability analysis for the Radiative Rayleigh-Taylor
  Instability} \label{sec:rrti}

We start from the equations of hydrodynamics with gravity and a
simple heating source term $H(X,T)$ in the total energy equation that depends on
temperature $T$ and composition $X$ to model radiation
\begin{eqnarray}\label{eq:euler}
 \frac{\partial \rho}{\partial t} + \vec{\nabla}\left(\rho\vec{u}\right) &=&
 0\cr
 \frac{\partial \rho \vec{u}}{\partial t} +
 \vec{\nabla}\left(\rho\vec{u}\otimes\vec{u} +
 P\right) &=& \rho \vec{g} \cr
 \frac{\partial \rho X}{\partial t} + \vec{\nabla}\left(\rho X\vec{u}\right) &=&
 0 \cr
 \frac{\partial \rho\mathcal{E}}{\partial t} +
 \vec{\nabla}\left(\vec{u}\left(\rho
 \mathcal{E}+P\right)\right) &=& \rho c_p H(X,T)
\end{eqnarray}
with the total energy $\mathcal{E}=e+u^2/2+\phi$, $e$ the internal energy,
$\vec{u}$ the velocity, $\phi$ the gravitational potential, the equation of
state (EOS) of an ideal gas $\rho e (\gamma-1) =P$, and the gravity
$\vec{g}=-\vec{\nabla}\phi$ (aligned with the y axis in the rest of the
paper). With the ideal gas law, the temperature and the 
(constant) mean molecular weight are linked by $P = \rho k_b T/\mu$.
We assume an hydrostatic background with
\begin{equation}\label{eq:stat}
  \frac{\partial P_0}{\partial y}= -\rho_0 g, \quad H(X_0,T_0)= 0
\end{equation}
assuming a discontinuity of composition in two parts of the domain with $X_0^+$
in the upper half and $X_0^-$ in the lower half and with a continuous density at the
interface $\rho_0$. We define $H_{T,X}$ 
the partial derivative of the source term with respect to temperature and
composition respectively and we assume for simplicity (and for the linear
stability analysis in Appendix~\ref{app:rrti}) that $H_{T,X}$ is constant at the
interface.

The classic Rayleigh-Taylor analysis predicts no instability
in the absence of density discontinuities. However, following
\citet{tremblin:2019}, we show in Appendix~\ref{app:rrti}
a linear stability analysis of the diabatic Rayleigh-Taylor instability in the
Boussinesq regime which gives a radiative Rayleigh-Taylor instability for the
system~\ref{eq:euler}:
\begin{equation}\label{eq:omega}
  \omega^2  = \frac{g k}{2} \frac{H_X}{T_0 H_T}\left( X_0^- - X_0^+\right)
\end{equation}
with $k$ the horizontal wavelength of the perturbation and $\omega$ the growth rate of the
instability. Equation \ref{eq:omega} is a simplified version of a more general
  diabatic Rayleigh-Taylor growth rate given in appendix (see
  \ref{eq:gr_dia}). The effect of a stable stratification in both side of the
  interface is neglected here but can be taken into account in the more general
  growth rate. We refer the reader to the appendix for more details.

We highlight that this growth rate is the discontinuous version of
the continuous case studied in \citet{tremblin:2019} with the
criterion for diabatic convection without mean-molecular-weight gradient nor
compositional source terms: $\nabla_X H_X >0$ with $\nabla_X=\partial \log
X/\partial \log P$. Qualitatively, the instability in the continuous and
discontinuous cases are easy to understand and rely on buoyancy: when a parcel of
fluid moves upward, if a compositional change during the motion induces heating,
the temperature increases and the density decreases which is destabilizing
relative to buoyancy. On the contrary if the compositional change induces
cooling, the temperature decreases and the density increases which is
stabilizing for an upward motion. To trigger the instability in a
  stratified medium, the impact of the source term has to induce heating for
  upward motions and cooling for downward motions and also be sufficiently large
  to overcome the stabilizing 
  effect of a stable stratification.

Remarkably, we get an instability without a discontinuity of density at
the interface, but with a heating source term that depends on composition
associated to a discontinuity of composition. This situation is typical of
cloud interfaces in planetary atmospheres, we will therefore explore this
instability with numerical simulations in the brown dwarf and rocky exoplanet
regimes.

\begin{figure}[btp]
\begin{centering}
\includegraphics[width=1.0\linewidth]{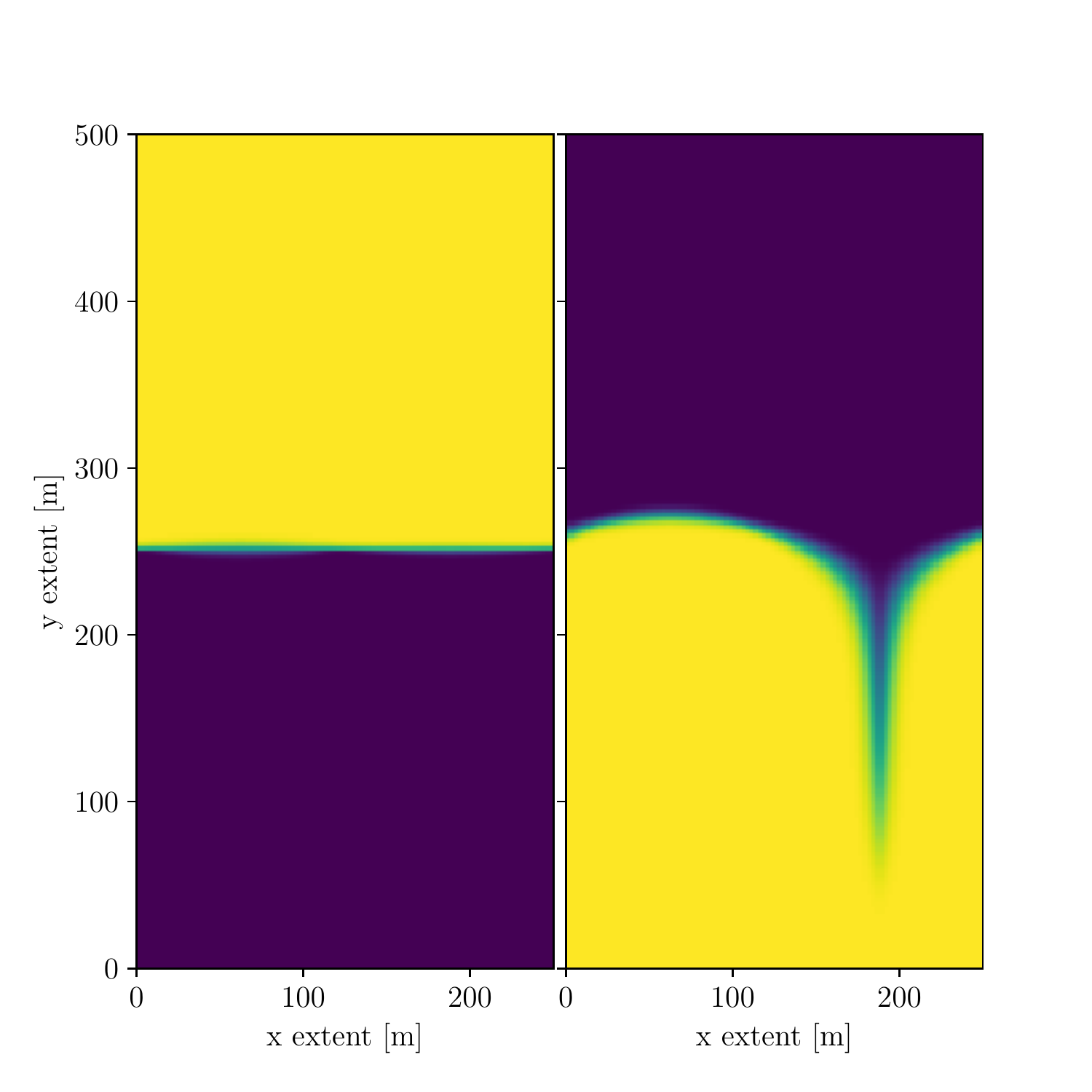}
\caption{Final 2D maps of the opacity tracer in a brown-dwarf regime with a
  negative vertical temperature gradient. Simulations are started from: left,
  a higher-opacity medium (yellow) on top of a lower-opacity medium (dark blue); right, a
  lower-opacity medium on top of an higher-opacity medium.} \label{fig:imshow_bd}
\end{centering}
\end{figure}

\begin{figure}[btp]
  \begin{centering}
    \includegraphics[width=1.0\linewidth]{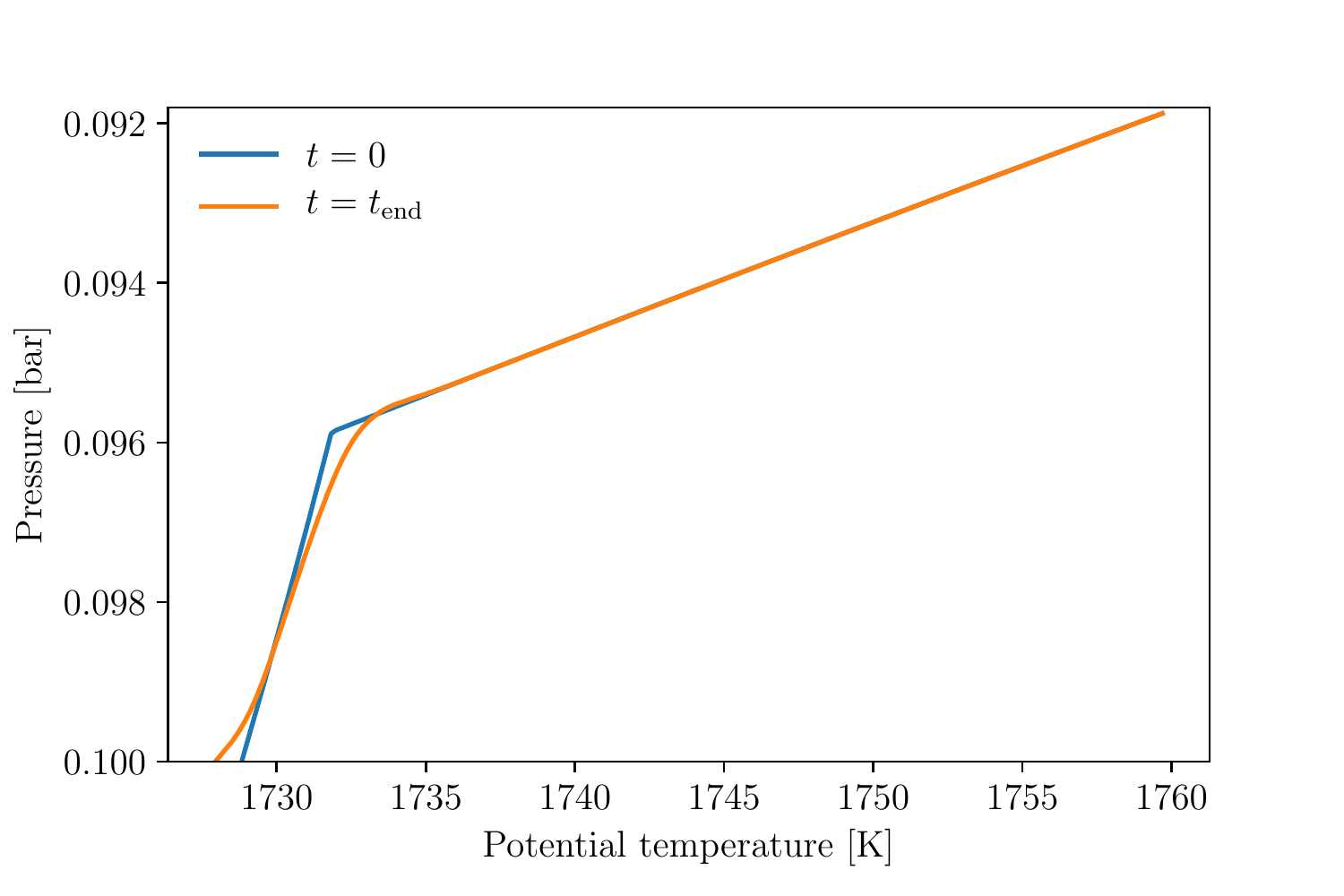}
    \includegraphics[width=1.0\linewidth]{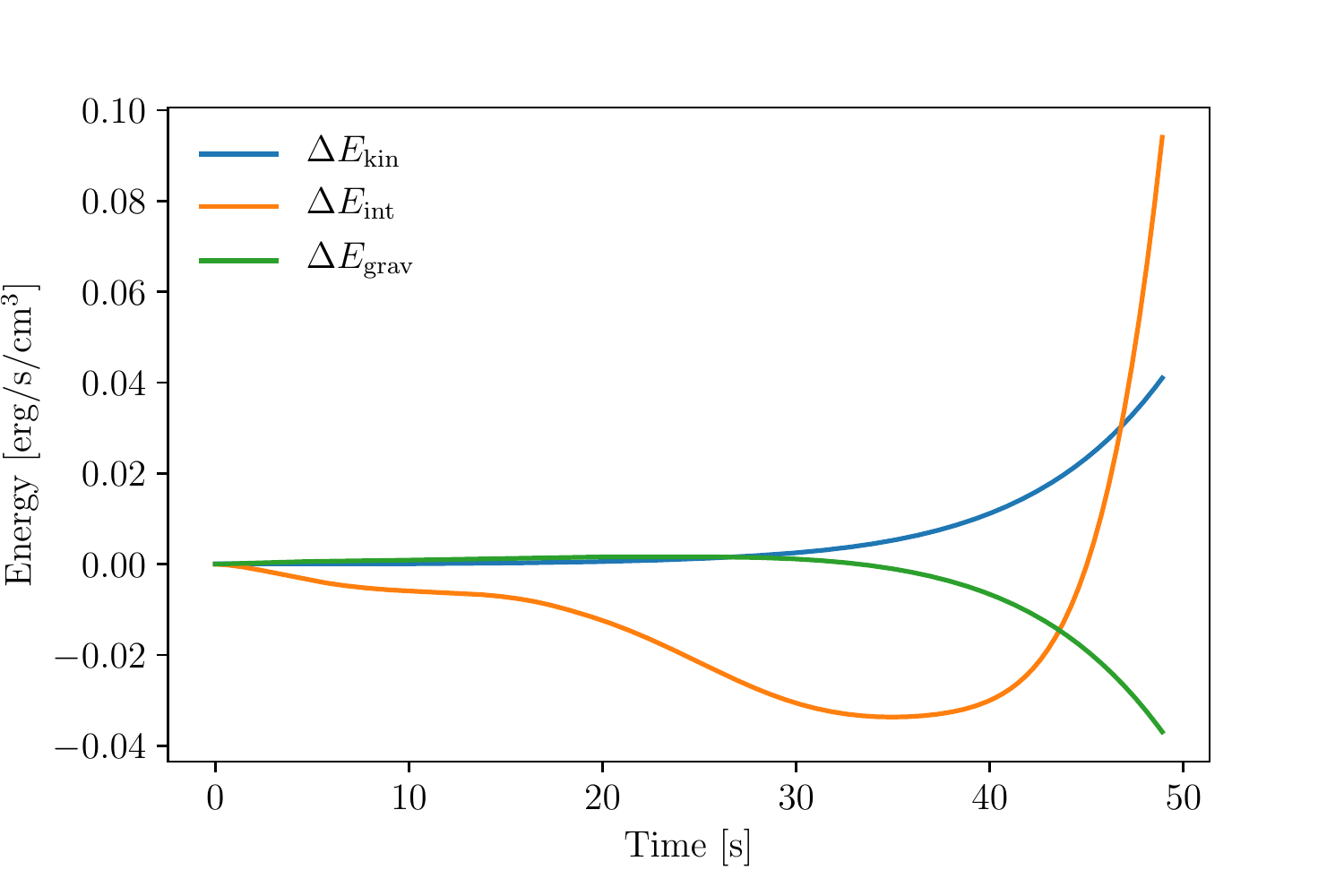}
\caption{Top: initial and final potential temperature profiles for the
  simulation with a lower-opacity medium on top of a higher-opacity one. Bottom:
evolution of the averaged kinetic, gravitational and internal
energies during the simulation. We plot the evolution of the differences from
their initial values.} \label{fig:diag}
\end{centering}
\end{figure}

\begin{figure}[btp]
\begin{centering}
\includegraphics[width=1.0\linewidth]{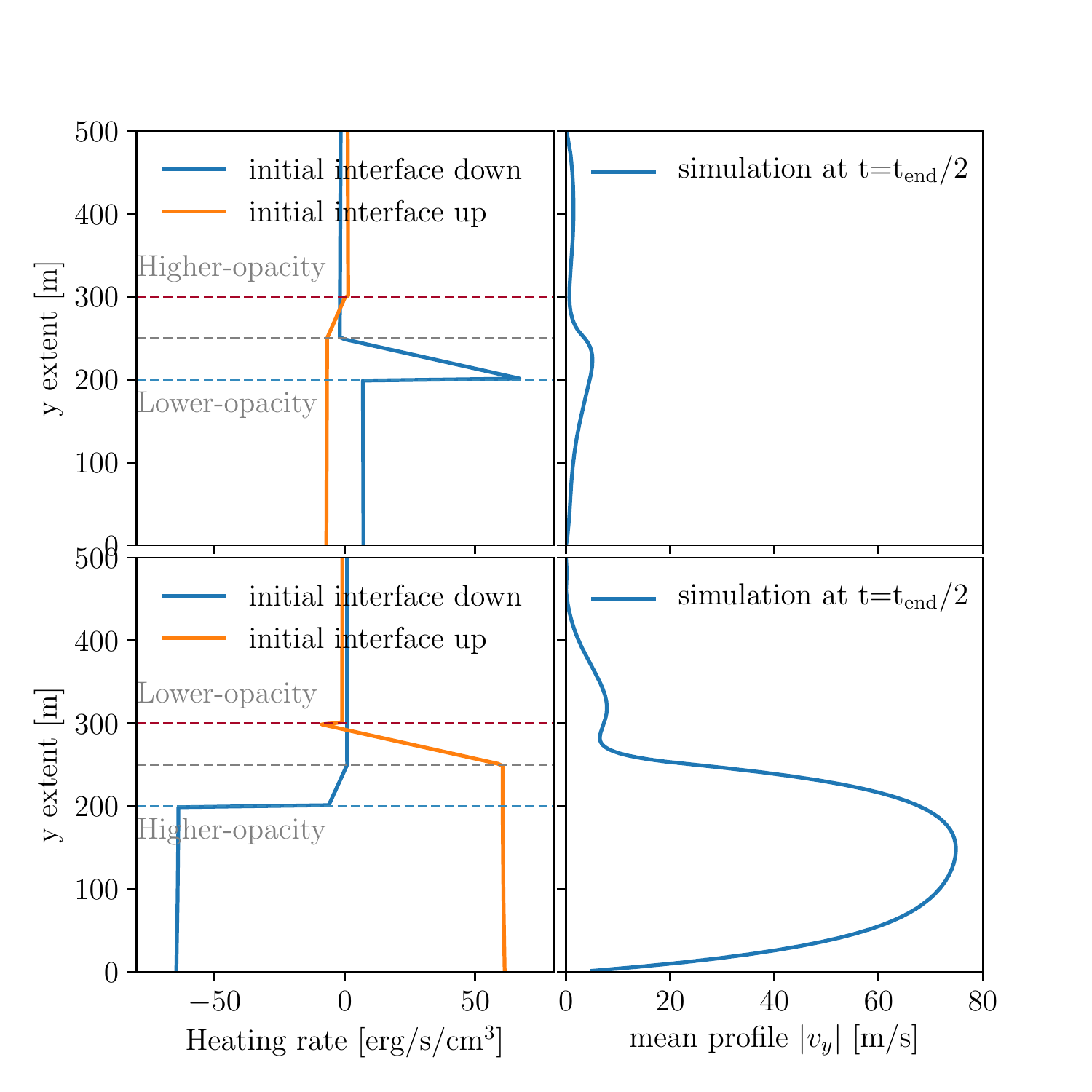}
\caption{Brown-dwarf regime with a negative vertical temperature gradient. Top,
  higher-opacity over lower-opacity; bottom: lower-opacity over higher-opacity. Left, 
  heating rate profiles when the position of the interface is displaced up or
  down by 10 \% from the initial condition. Right: mean vertical velocity
  profile in the simulation at $t=t_\mathrm{end}/2$.} \label{fig:profiles_bd}
\end{centering}
\end{figure}

\section{Numerical method and setup} \label{sec:model}

We use the code \texttt{ARK} \citep{padioleau:2019} to solve the hydrodynamics
equations in two dimensions (x and y) with a 
radiative source term as presented in \citet{tremblin:2019}. In this paper, we
ignore changes in the mean molecular weight, keep it constant, and trace with a
scalar the opacity of the medium. The equations solved in this setup are the
system~\ref{eq:euler} with
\begin{equation}\label{eq:heating}
 H(X,T) = \frac{4\pi \kappa(X)}{c_p} (J-\sigma T^4/\pi)
 \end{equation}
which is equivalent to the divergence of the radiative flux
  \citep[see][]{mihalas:1984}. the The opacity
is $\kappa(X) = X \kappa_1 + (1-X) \kappa_2$ tracing gases with two different
opacities $\kappa_1$ and $\kappa_2$. The mean grey intensity is
$J=(I^\uparrow+I^\downarrow)/2$ with $I^\uparrow$ and $I^\downarrow$ the upward
and downward intensities computed using the radiative transfer equation in a
standard two-stream approximation \citep[see][for details]{tremblin:2019}. We
treat here only absorption opacities and scattering is ignored. 

The hydrodynamics solver is a well-balanced and all-regime solver
extensively described and tested in \citet{padioleau:2019}. These properties of
the numerical method are essential for this study:

\begin{itemize}
  \item the well-balanced property means that the method is able to maintain the
    hydrostatic balance at machine precision. This gives us a very
    precise control of the instability: even when initialized on an unstable
    equilibrium, the simulation does not develop velocities unless we
    explicitely add a perturbation.
  \item the all-regime property means that the solver has a low-mach correction
    to reach a high accuracy in the regime of low velocities. This low-mach
    correction is activated in all the simulations presented in this paper and
    is also essential to capture the instability in the regime we have explored.
\end{itemize}

There is no diffusivity, viscosity, and sub-grid turbulence in the
  model. The dissipation relies only on the numerical diffusion at the grid
  scale of the all-regime numerical method. Thanks to the low-Mach correction,
  that dissipation is significantly reduced, as shown in \citet{padioleau:2019}.

The initial conditions are all chosen in a similar way: we initialize the scalar
tracing opacities $X=X_0$ to 0 and 1 or vice versa in the upper and lower half
of the domain. We then pre-compute a pressure/temperature profile that satisfy
the discrete version of the hydrostatic balance and energy conservation in
Eq.~\ref{eq:stat} starting from an imposed base pressure $P_\mathrm{bot}$. As
explained above with 
the well-balanced property, the simulation will remain static in the absence of
perturbations even if these initial conditions are unstable.

The boundary conditions for the density and pressure are imposed by a linear
extrapolation of the temperature and an extrapolation of the hydrodystatic
balance at the top and bottom boundaries of the domain in the y direction. We also impose
zero velocities. For the radiative variables, we impose the downward radiative
flux at the top $\pi I^\downarrow_\mathrm{top}$ and the net radiative flux at
the bottom $\pi(I^\uparrow_\mathrm{bot}-I^\downarrow_\mathrm{bot})$. The
  downward radiative flux at the top represents the emission of the layers of
  the atmosphere that are above our computational domain. Imposing
the net radiative flux at the bottom implies that the downward intensity at the
bottom is re-emitted upward by deeper layers or a planetary surface. The
magnitude of these fluxes are chosen in conjonction with the grey opacities
$\kappa_1$ and $\kappa_2$ to ensure a realistic
temperature profile, furthermore the adiabatic index $\gamma$ is adjusted such that this
idealized setup is stable to Schwarschild convection and allow the study of
other instabilities. The boundary conditions in the x direction are periodic.

The initial perturbation is adapted differently in the brown-dwarf and
earth-like regimes, we detail this
choice in Sect.~\ref{sec:bd-model} and Sect.~\ref{sec:earth-model}. 

\begin{table}
  \centering
  \caption{Parameters used for the simulations of an opacity interface in a
  brown-dwarf regime}\label{tab:param-bd}
\begin{tabular}{l|l}
  x$_\mathrm{max}$, y$_\mathrm{max}$ [m] & 250, 500 \\
  nx, ny & 100, 200 \\
  t$_\mathrm{end}$ [s] & 60 \\
  $\log(g)$ & 5 \\
  P$_\mathrm{bot}$ [bar] & 0.1 \\
  $\gamma$ & 1.7 \\
  $\mu$ & 2.4 \\
  $\pi I^\downarrow_\mathrm{top}$ [erg/s/cm$^2]$& 2.87$\times$10$^8$
  (T$_\mathrm{rad}$=1500 K) \\
  $\pi(I^\uparrow_\mathrm{bot} -I^\downarrow_\mathrm{bot})$ [erg/s/cm$^2$]&
  3.72$\times$10$^8$ (T$_\mathrm{rad}$=1600 K) \\
  $\kappa_1$ [cm$^2$/g] & 0.2 \\
  $\kappa_2$ [cm$^2$/g] & 2.0 \\
  A (see eq.~\ref{eq:pert}) & 10$^{-4}$ \\
  m (see eq.~\ref{eq:pert}) & 2 \\
  w (see eq.~\ref{eq:pert}) & 0.25 \\
\end{tabular}
\end{table}

\begin{figure}[btp]
\begin{centering}
\includegraphics[width=1.0\linewidth]{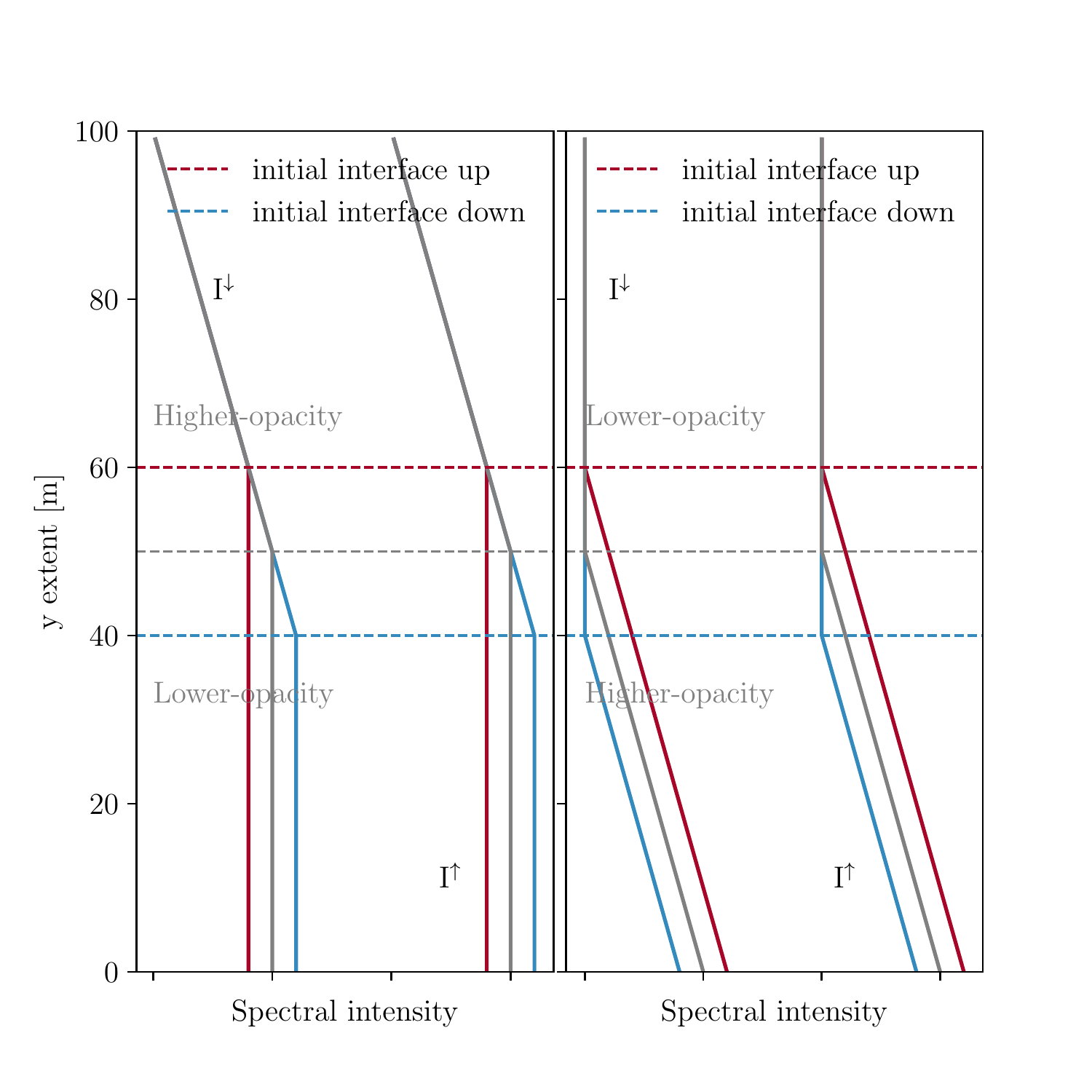}
\caption{Simplified downward and upward intensity profiles in an atmosphere with
a negative vertical gradient of temperature. Grey profiles are profiles for an
initial condition at radiative equilibrium with the interface at the middle of
the box. Red and blue profiles corresponds to intensities when the interface has
been moved up and down.} \label{fig:intensity_Tdec}
\end{centering}
\end{figure}

\begin{figure}[btp]
\begin{centering}
\includegraphics[width=1.0\linewidth]{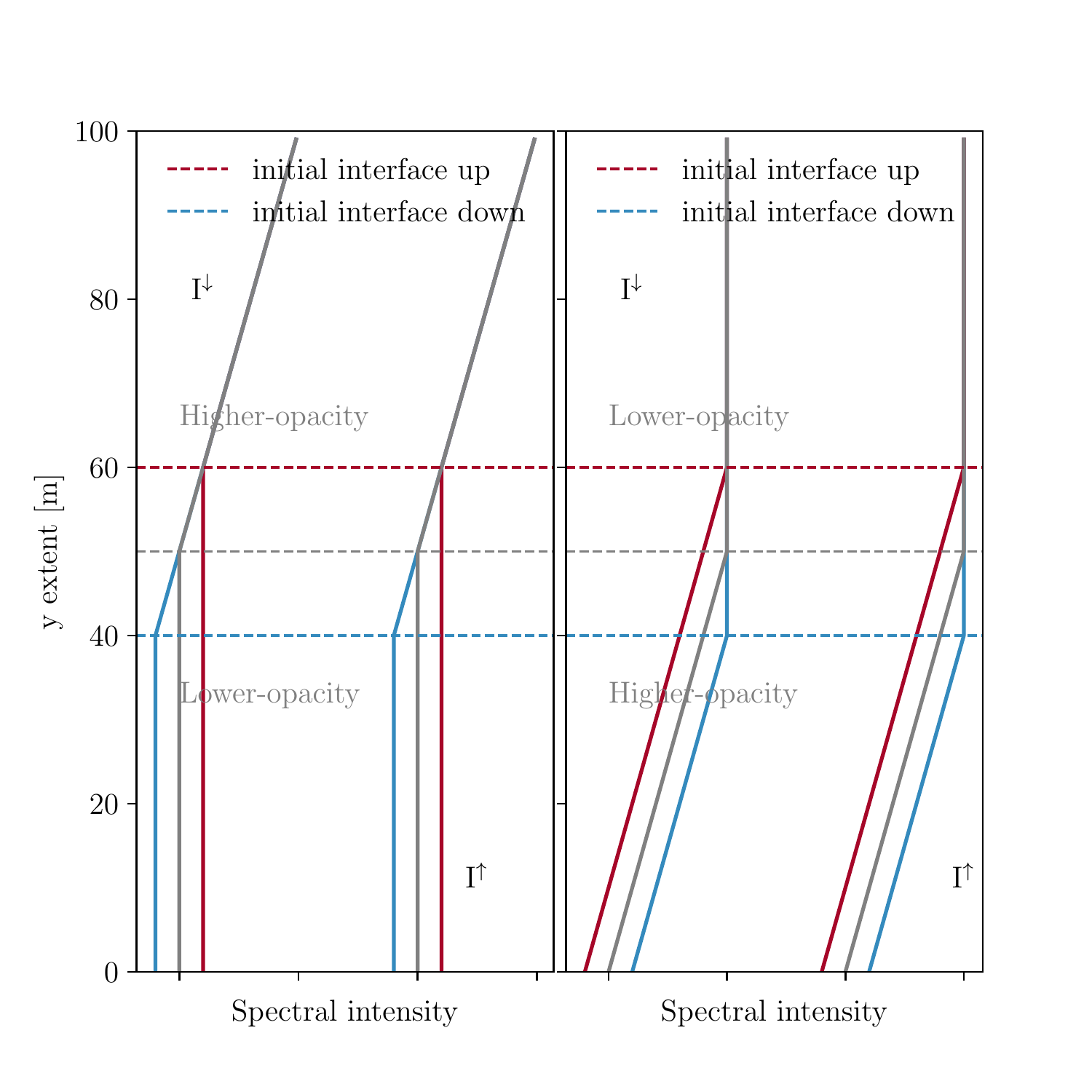}
\caption{Simplified downward and upward intensity profiles in an atmosphere with
a positive vertical gradient of temperature. Grey profiles are profiles for an
initial condition at radiative equilibrium with the interface at the middle of
the box. Red and blue profiles corresponds to intensities when the interface has
been moved up and down.} \label{fig:intensity_Tinc}
\end{centering}
\end{figure}

\section{RRTI in the Brown-dwarf regime} \label{sec:bd-model}

We first choose our parameters to model a stably stratified part of the upper
atmosphere of a L dwarf with effective temperature $T_\mathrm{eff}\sim 1600 K$
and surface gravity $\log(g)=5$. All the parameters including the opacities and
radiative boundary conditions are given in Tab.~\ref{tab:param-bd} and are set to
reproduce a temperature gradient decreasing with height stable to Schwarzschild
convection. We assume the presence of an opacity interface e.g. coming from
condensation of silicates or iron, with an opacity ratio of a factor 10.
We use a perturbation in the initial vertical velocity:

\begin{equation}\label{eq:pert}
u_{0,y}(x,y)=A c_s \sin(m\pi x/x_\mathrm{max})e^{-(y-0.5y_\mathrm{max})^2/(w y_\mathrm{max})^2}
 \end{equation}
 with $c_s$ the local sound speed, $x_\mathrm{max}$, $y_\mathrm{max}$ the
 horizontal and vertical extent of the box, respectively.

Figure~\ref{fig:imshow_bd} shows the maps of the opacity tracer at
$t=t_\mathrm{end}$ when we start from the higher-opacity medium on top of the
lower-opacity one (left panel) and vice versa (right panel) and clearly
demonstrate that higher-opacity over 
lower-opacity is stable while lower-opacity over higher-opacity is unstable. We highlight
that higher-opacity versus lower-opacity only refers to $\kappa_2$ versus $\kappa_1$
and not to an optically thick versus thin transition: in this setup, both medium
are optically thin, the total optical depth after crossing both media is
0.18. This instability can be directly linked to the dependance of 
the opacity to a local tracer given in Sect.~\ref{sec:rrti}.

We show in the top panel of Fig.~\ref{fig:diag}, the initial and final profile
of potential temperature $\theta=T(P_\mathrm{ref}/P)^{(\gamma-1)/\gamma}$ in the
unstable simulation. It shows that the potential temperature is always
increasing with height, hence the simulations is always stably stratified
i.e. stable to thermal convection. The bottom panel shows the evolution of the
averaged internal, gravitational and kinetic energy in the simulation. One can
see that the kinetic energy is equal to the opposite of the gravitational
energy: this is expected for a buoyancy instability that convert gravitational
potential energy into kinetic energy. The evolution of the internal energy is
not conservative which is expected because of the radiative transfer source
term. Its evolution leads to conditions prone to the RRTI instability which we
will explore in more details below.

In order to explore the link to RRTI, we show in
Fig.~\ref{fig:profiles_bd}, the heating rate computed from our initial conditions
when the interface is artificially displaced by 10 \% up or down in the
box. Since the heating rate is exactly zero when the interface is at the middle
of the box, this artificial displacement allows us to probe numerically the
evolution of $H$ with composition and get an estimate of $H_X$ in our simulations.
As explained above, we can expect the flow to be unstable to buoyancy when an
upward displacement 
leads to heating and a downward displacement to cooling, while upward
displacements inducing cooling would correspond to a stable situation. Indeed
figure~\ref{fig:profiles_bd} shows that vertical velocities and 
the instability appear when the displacement of the interface is unstable to
buoyancy. We can
provide a theoretical estimate of the growth rate using Eq.~\ref{eq:omega} and
by computing numerically $H_{X,T}$ at the interface from the initial
conditions. Such an estimate gives a growth rate of 0.04 s$^{-1}$ while the
measured growth rate in the simulation is 0.07 s$^{-1}$ (70 \% higher, measured
between t=10 and t=50 s before reaching saturation). The
theoretical estimate provides the correct order of magnitude, and the agreement
can be considered as relatively good given the approximations in the linear
stability analysis that partly ignore the complexity and non-locality of radiative
transfer.

Figure~\ref{fig:profiles_bd} also shows that the displacement of the interface
leads to a non-local heating and cooling preferentially in the higher-opacity
medium. Such a non-locality is not surprising: when the
medium is optically thin, a local modification of its opacity has an impact in the entire
domain. Heating and cooling happens preferentially in the higher-opacity domain simply
because the heating rate is proportional to the opacity (see Eq.~\ref{eq:heating}).
The mean velocity profile shows that motions are indeed appearing preferentially
in the entire higher-opacity medium and not only at the interface. This non-locality and
asymmetry between the two media can explain the rapidly non-linear and strongly
asymetric development of the perturbations in Fig.~\ref{fig:imshow_bd}: the
downward column is narrow and fast while the upward motions are spatially large
and slow.

In order to study this radiative heating and cooling more in depth, we need to
explore the evolution of the upward and downward intensities as a function of
the interface displacement. We simplify the problem by assuming that the
radiative transfer equation leads to a simplified solution of the type $I(y) =
e^{-\kappa\rho \Delta y} I(y_0) + (1-e^{-\kappa \rho \Delta y}) \sigma
T(y)^4/\pi$ with a vanishing
opacity in the lower-opacity medium i.e. $I(y)=I(y_0)$ and a small opacity in the
higher-opacity medium such that $I(y)$ is in the linear regime as a function of $y$.
At equilibrium in the initial condition, the radiative flux is constant and the
heating rate is zero, this implies that $I^\uparrow-I^\downarrow$ is constant
and $(I^\uparrow+I^\downarrow)/2=\sigma T^4/\pi$, i.e. the intensity
profiles increase or decrease with altitude similarly to temperature. Under
these assumptions, Fig.~\ref{fig:intensity_Tdec} shows the typical equilibrium
downward and upward intensity profiles in grey when the temperature is
decreasing with altitude. We detail the two different cases:

\begin{itemize}
\item higher-opacity over lower-opacity: Starting from the top boundary, the downward intensity
is increasing when going down, which corresponds to emission in the higher-opacity
medium, it is then constant when propagating through the lower-opacity
medium. Then starting from the bottom boundary the upward intensity is constant
through the lower-opacity medium, its value being fixed by the net radiative flux
boundary condition, and then decrease in the higher-opacity medium which
correponds to absorption. When the interface is moving up (red curves), the optical path in
the higher-opacity medium is decreasing, corresponding to a decrease of emission
for the downward photons which decreases first the downard intensity and then
the upward intensity because of the imposed net radiative flux at the bottom
boundary. The decrease of both intensities hence results in a decrease of
  the mean grey intensity J. Using Eq.(\ref{eq:heating}) together with the fact that $T$ is
  constant, means there is a net cooling of the atmosphere when the
  interface is moving up, which is stabilizing.
\item lower-opacity over higher-opacity: Starting from the top boundary, the downward
  intensity is constant when going down in the lower-opacity medium, then
  increases in the higher-opacity medium, which corresponds to emission in the higher-opacity
  medium. Then strating from the bottom boundary the upward intensity is
  decreasing in the higher-opacity medium which corresponds to absorption and is
  constant through the lower-opacity medium. When the interface is moving up, the
  optical path in the higher-opacity medium is increasing, corresponding to an increase
  of emission for the downward photons which increases first the downward
  intensity and then the upward intensity because of the imposed net radiative
  flux at the bottom boundary. The increase of both intensities hence
    results in an increase of
  the mean grey intensity J. Using Eq.(\ref{eq:heating}) together with the fact that $T$ is
  constant, means there is a net heating of the atmosphere when the
  interface is moving up, which is destabilizing. 
\end{itemize}
To summarize, when the temperature is decreasing with altitude, a decrease of
the optical path in the higher-opacity medium results in a decrease in emission hence
radiative cooling and an increase of the optical path results in an increase in
emission hence radiative heating. Interestingly, it can be expected from
Fig.~\ref{fig:intensity_Tdec} that this behavior is inverted in the presence of
a temperature inversion. Fig.~\ref{fig:intensity_Tinc} shows the expected
intensity profiles if the temperature is increasing with height. The same
analysis can be done but now, the higher-opacity medium is absorbing the downward
photons (since the intensity is decreasing along the path). Consequently a
decrease of the optical path in the higher-opacity medium 
results in a decrease in absorption hence radiative heating while an increase of
the optical path in the higher-opacity medium results in an increase in absorption hence
radiative cooling. It is then clear that we should expect the instability to be
inverted in that case: when the temperature is increasing with altitude, a
higher-opacity medium on top of a lower-opacity medium should be unstable and lower-opacity
over higher-opacity should be stable. We explore this possibility in an earth-like
regime in the next section.

\begin{table}
  \centering
  \caption{Parameters used for the simulations of an opacity interface in a
  earth-like regime with a negative vertical gradient of
  temperature}\label{tab:param-earth-Tdec}
\begin{tabular}{l|l}
  x$_\mathrm{max}$, y$_\mathrm{max}$ [m] & 50, 100 \\
  nx, ny & 100, 200 \\
  t$_\mathrm{end}$ [s] & 1260 \\
  logg & 2.99 \\
  P$_\mathrm{bot}$ [bar] & 0.8 \\
  $\gamma$ & 1.5 \\
  $\mu$ & 28.97 \\
  $\pi I^\downarrow_\mathrm{top}$ [erg/s/cm$^2]$& 3.24$\times$10$^5$
  (T$_\mathrm{rad}$=275 K) \\
  $\pi(I^\uparrow_\mathrm{bot} -I^\downarrow_\mathrm{bot})$ [erg/s/cm$^2$]&
  5.19$\times$10$^2$ (T$_\mathrm{rad}$=55 K) \\
  $\kappa_1$ [cm$^2$/g] & 10$^{-3}$ \\
  $\kappa_2$ [cm$^2$/g] & 1.0 \\
  $\tau_\mathrm{forcing}$ [s] & $\infty$ \\
  A (see eq.~\ref{eq:pert_earth}) & 2$\times$10$^{-2}$ \\
  m (see eq.~\ref{eq:pert_earth}) & 2 \\
\end{tabular}
\end{table}

\begin{figure}[btp]
\begin{centering}
\includegraphics[width=1.0\linewidth]{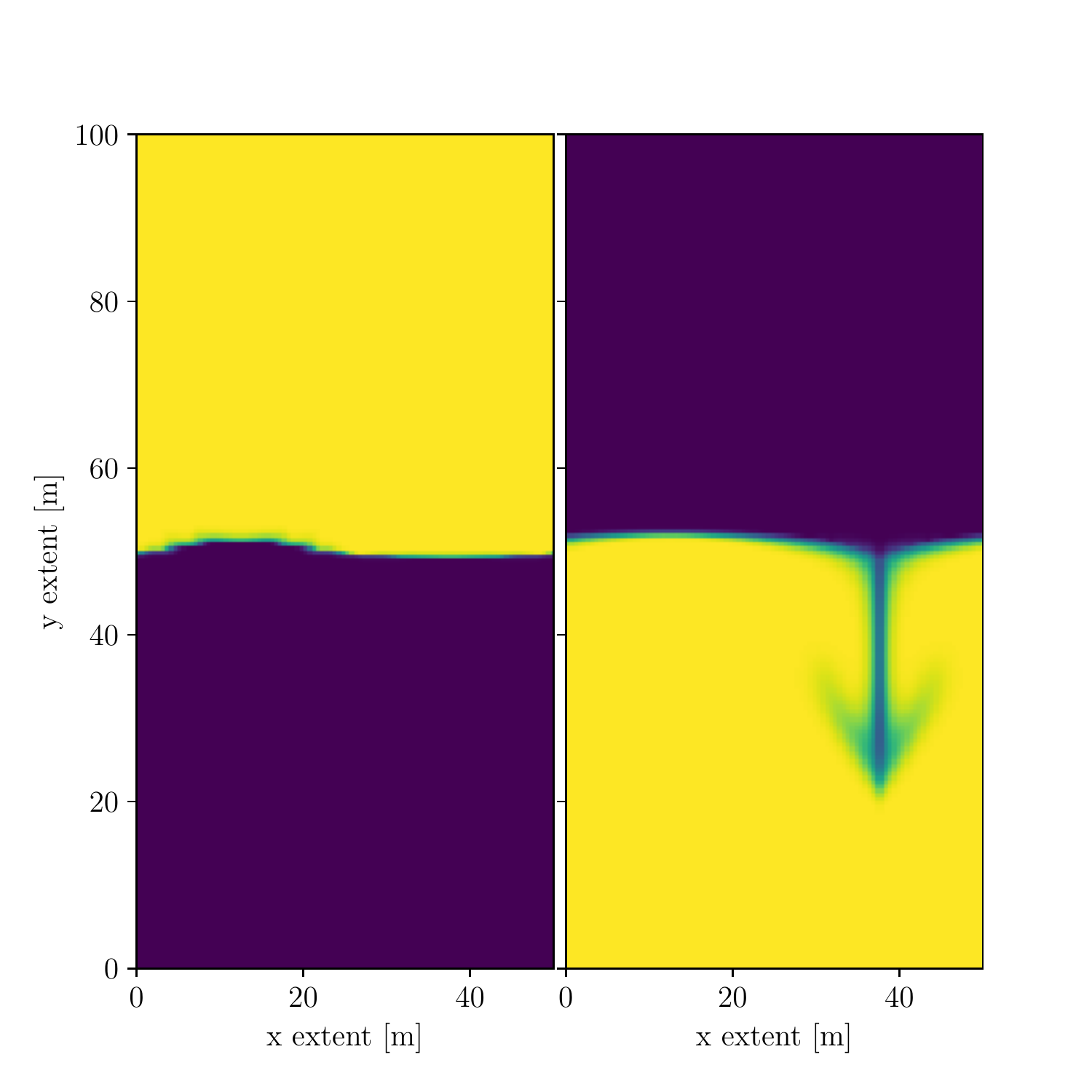}
\caption{Final 2D maps of the opacity tracer in an earth-like regime with
  a negative vertical temperature gradient. Simulations are started from: left,
  a higher-opacity medium (yellow) on top of a lower-opacity medium (dark blue); right, a
  lower-opacity medium on top of a higher-opacity medium.} \label{fig:imshow_earth_Tdec}
\end{centering}
\end{figure}

\begin{figure}[btp]
\begin{centering}
\includegraphics[width=1.0\linewidth]{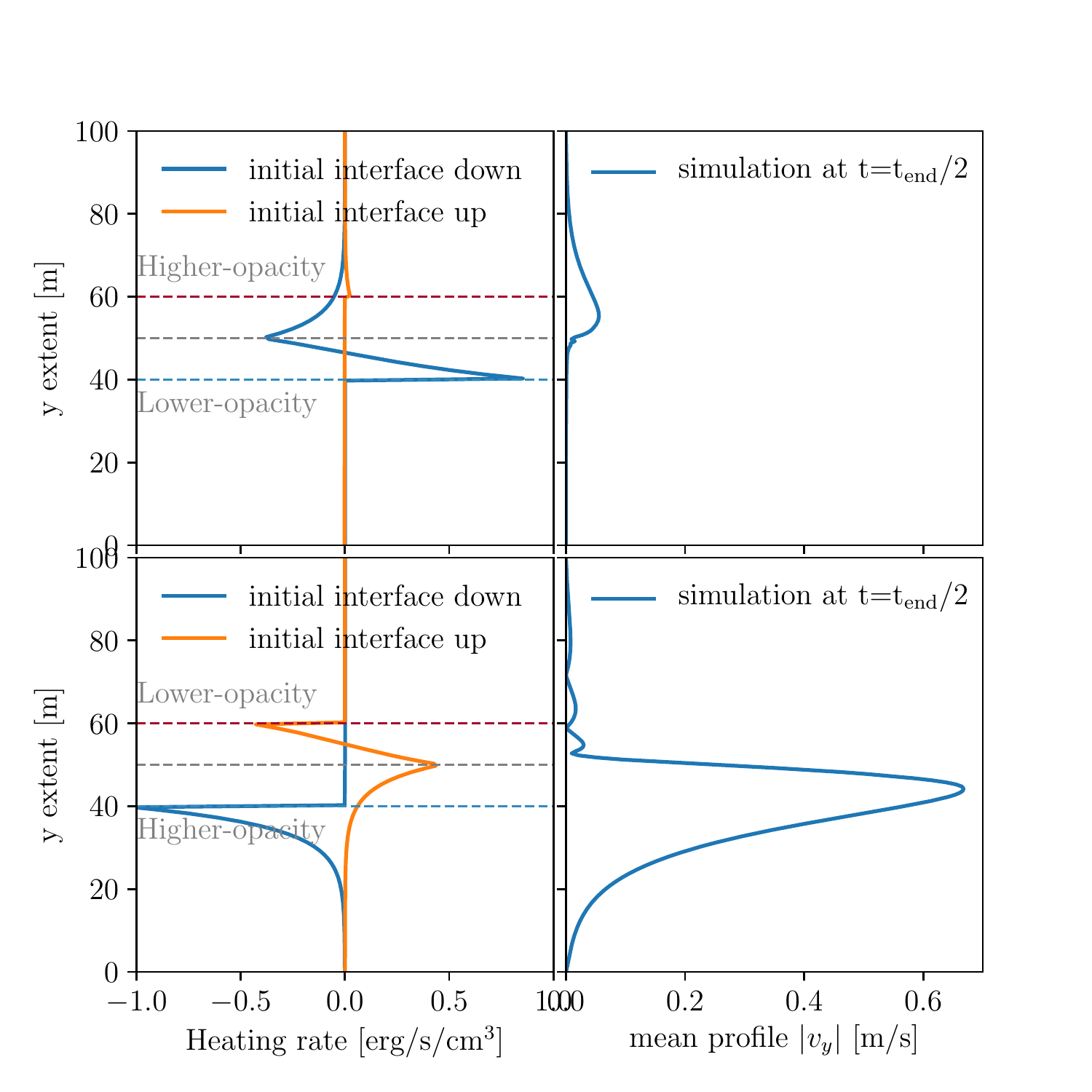}
\caption{Earth-like regime with a negative vertical temperature gradient. Top,
  higher-opacity over lower-opacity; bottom: lower-opacity over higher-opacity. Left,
  heating rate profiles when the position of the interface is displaced up or
  down by 10 \% from the initial condition. Right: mean vertical velocity
  profile in the simulation at $t=t_\mathrm{end}/2$.} \label{fig:profiles_earth_Tdec}
\end{centering}
\end{figure}

\begin{table}
  \centering
  \caption{Forcing parameters used for the simulations of an opacity interface in
  a earth-like regime with a positive vertical gradient of temperature. All the
other parameters are similar to
Tab.~\ref{tab:param-earth-Tdec}.} \label{tab:param-earth-Tinc} 
\begin{tabular}{l|l}
  T$_\mathrm{forcing,top}$ [K] & 275 \\
  T$_\mathrm{forcing,bot}$ [K] & 270 \\
  $\tau_\mathrm{ref}$ [s]      & 875 \\
  $\kappa_\mathrm{ref}$ [cm$^2$/g] & 1.0 \\
  t$_\mathrm{end}$ [s] & 4500 \\
\end{tabular}
\end{table}

\begin{figure}[btp]
\begin{centering}
\includegraphics[width=1.0\linewidth]{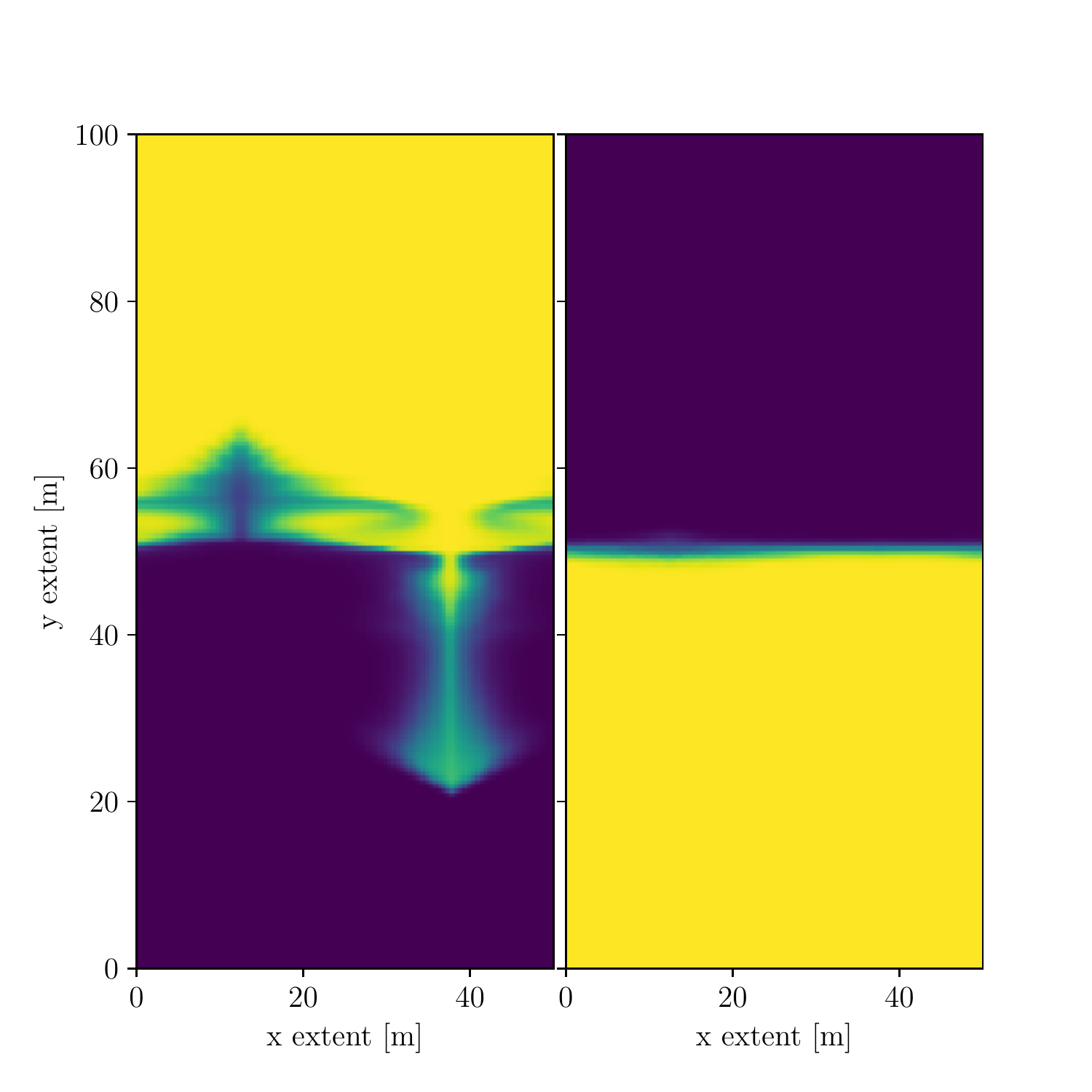}
\caption{Final 2D maps of the opacity tracer in an earth-like regime with
  a positive vertical temperature gradient. simulations are started from: left,
  a higher-opacity medium (yellow) on top of a lower-opacity medium (dark blue); right, a
  lower-opacity medium on top of a higher-opacity medium.} \label{fig:imshow_earth_Tinc}
\end{centering}
\end{figure}

\begin{figure}[btp]
\begin{centering}
\includegraphics[width=1.0\linewidth]{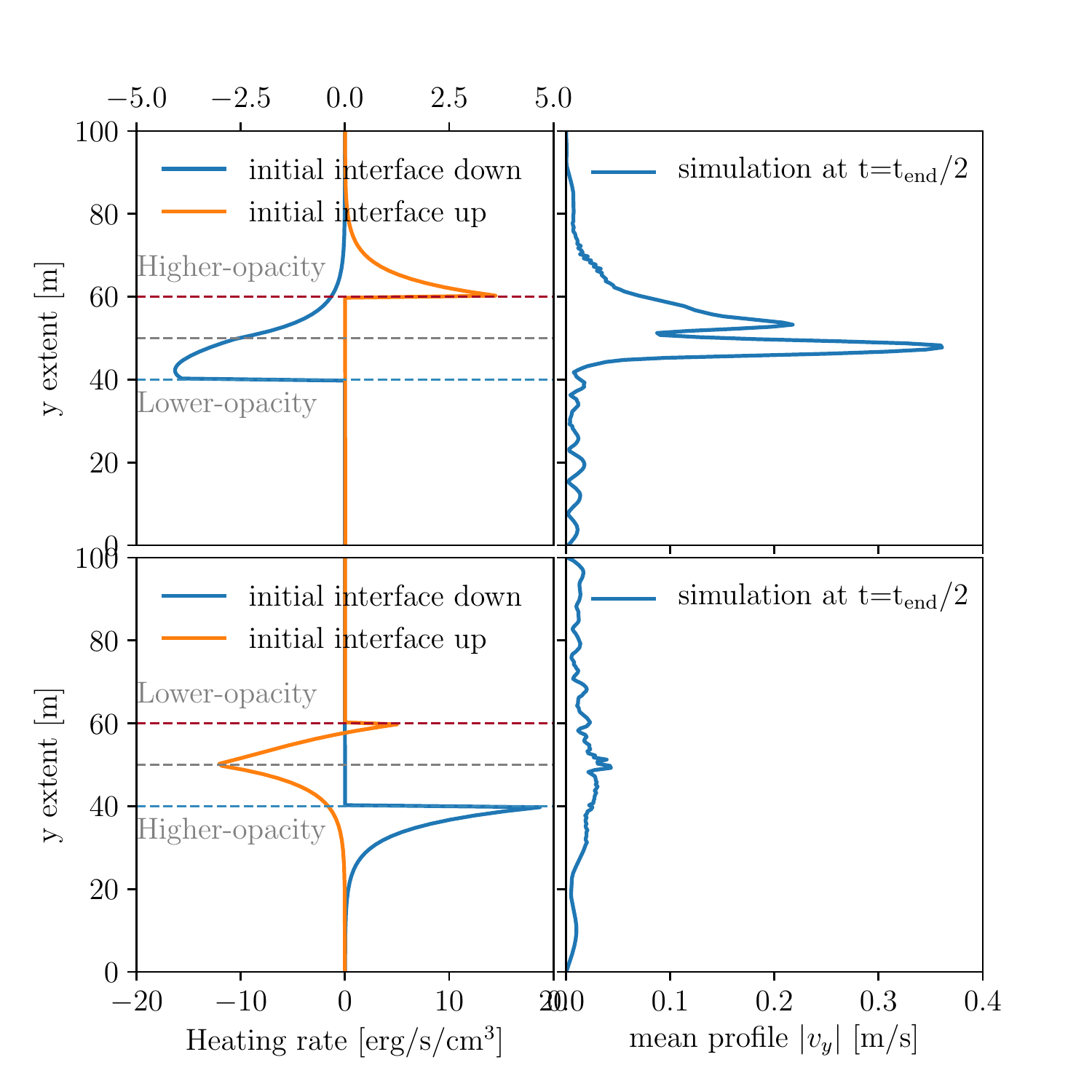}
\caption{Earth-like regime with a positive vertical temperature gradient. Top,
  higher-opacity over lower-opacity; bottom: lower-opacity over higher-opacity. Left,
  heating rate profiles when the position of the interface is displaced up or
  down by 10 \% from the initial condition. Right: mean vertical velocity
  profile in the simulation at $t=t_\mathrm{end}/2$.} \label{fig:profiles_earth_Tinc}
\end{centering}
\end{figure}

\section{Earth-like regime: Dependance on the temperature gradient}\label{sec:earth-model}

In order to explore the behavior of the RRTI with a temperature
inversion, we perform simulations in an earth-like regime for which a positive
vertical gradients of temperature could arise either because of irradiation
condensation/evaporation or
hydrodynamical effects. We therefore need to impose a temperature inversion by
adding a forcing term in the heating source term
\begin{equation}\label{eq:heating2}
 H(X,T) = \frac{4\pi \kappa(X)}{c_p} (J-\sigma T^4/\pi)-
 \frac{T-T_\mathrm{forcing}}{\tau_\mathrm{forcing}}
 \end{equation}
with $T_\mathrm{forcing}$ an imposed profile with a positive vertical gradient
and $\tau_\mathrm{forcing}$ the timescale for the forcing. Before that, we
first consider a purely radiative setup
(i.e. $\tau_\mathrm{forcing}\rightarrow \infty$)
and adjust the parameters (see Tab.~\ref{tab:param-earth-Tdec}) to reproduce the
unstable behavior in a negative vertical gradient of temperature (see
Fig.~\ref{fig:imshow_earth_Tdec}). The radiative balance in Earth atmosphere is
typically in the infrared at wavelength around 10 $\mu$m. We assume a ratio of
opacity of 1000 for Earth clouds \citep{kokhanovsky:2004} and adjust the
radiative boundary conditions to get a temperature/pressure profile that matches
the international standard atmosphere (ISA) values at 2~km altitude and is stable
to Schwarschild convection. The instability appears to be weaker in the
earth-like regime than in the brown-dwarf regime, a velocity perturbation in
that context tends to trigger sound waves that are reflected on the boundaries
of the domain and strongly interfere with the interface. To overcome this
numerical limitation, we have used an interface perturbation rather than a
velocity perturbation:
\begin{equation}\label{eq:pert_earth}
y_\mathrm{int}(x)= 0.5 y_\mathrm{max}(1+A \sin(m\pi x/x_\mathrm{max}))
\end{equation}
This type of perturbation appears to have a much weaker interaction with the boundaries of
the simulation. Figure~\ref{fig:imshow_earth_Tdec} shows that we recover the
behavior expected from the brown-dwarf setup: higher-opacity over lower-opacity is stable
and lower-opacity over higher-opacity is unstable. Figure~\ref{fig:profiles_earth_Tdec}
shows the behavior of the heating rate when we displace the interface in the
initial conditions and the velocity profile at the middle of the simulation. The
heating rate behaves similarly to the brown-dwarf regime: in the
lower-opacity over higher-opacity case, an upward displacement of the interface results in
radiative heating and a downward displacement to radiative cooling, hence the
interface is unstable to buoyancy. The difference with the brown-dwarf regime is
that the opacity jump is larger and the higher-opacity medium is optically thick. Indeed
we can also see in Fig.~\ref{fig:profiles_earth_Tdec} (left) that heating and cooling are
localized close to the interface which is a consequence of optically-thick
radiative transfer that degenerates to local thermal diffusion. The right panel
also shows that the motions tend to be localized closer
to the interface. This difference does not impact the general behavior of the
instability.

In order to explore the behavior of the instability with a temperature
inversion, we use a simple newtonian forcing as proposed in
Eq.~\ref{eq:heating2}. This newtonian cooling is a very simplified model to force
a temperature inversion that could arise from irradiation, phase change or
dynamics. The initial profile is computed to be at equilibrium $H=0$ for
  Eq.~\ref{eq:heating2}, i.e. including the Newtonian cooling. We point out that
  the argument made in Sect.\ref{sec:bd-model} to
  explain the instability in the presence of a temperature inversion still apply
  here because Newtonian cooling depends only on temperature.
We use a linear forcing profile with altitude and
parametrize the top and bottom forcing temperature (see
Tab.~\ref{tab:param-earth-Tinc}). We also choose a forcing 
timescale of the form $\tau_\mathrm{forcing} =
\tau_\mathrm{ref}\times(\kappa_\mathrm{ref}/\kappa(X))$. This form allows to
invert the higher-opacity and lower-opacity medium in the simulation and keep a
continuous temperature profile stable to Schwarzschild convection. It is also reasonable
to assume that irradiating and dynamical forcing will be more efficient in the
higher-opacity medium. All the parameters for the forcing are listed in
Tab.~\ref{tab:param-earth-Tinc} and we show in Fig.~\ref{fig:imshow_earth_Tinc}
the opacity tracer at the end of the simulation. As expected from
Fig.~\ref{fig:intensity_Tinc}, the instability is inverted: higher-opacity medium over
lower-opacity medium is unstable and lower over higher-opacity is stable when the
temperature is increasing with altitude. We also show in
Fig.~\ref{fig:profiles_earth_Tinc} the profiles of the heating rate when the
interface is displaced upward and downward. It shows that in the
higher-opacity over lower-opacity case, an upward displacement results in radiative
heating and a downward displacement in radiative cooling and confirm the
instability regime. As in Fig.~\ref{fig:profiles_earth_Tdec}, we can also see
that heating and cooling are localized close to the interface because of the
optically thick regime in the higher-opacity medium similarly to the motions triggered
by the instability.

\begin{figure}[btp]
\begin{centering}
\includegraphics[width=1.0\linewidth]{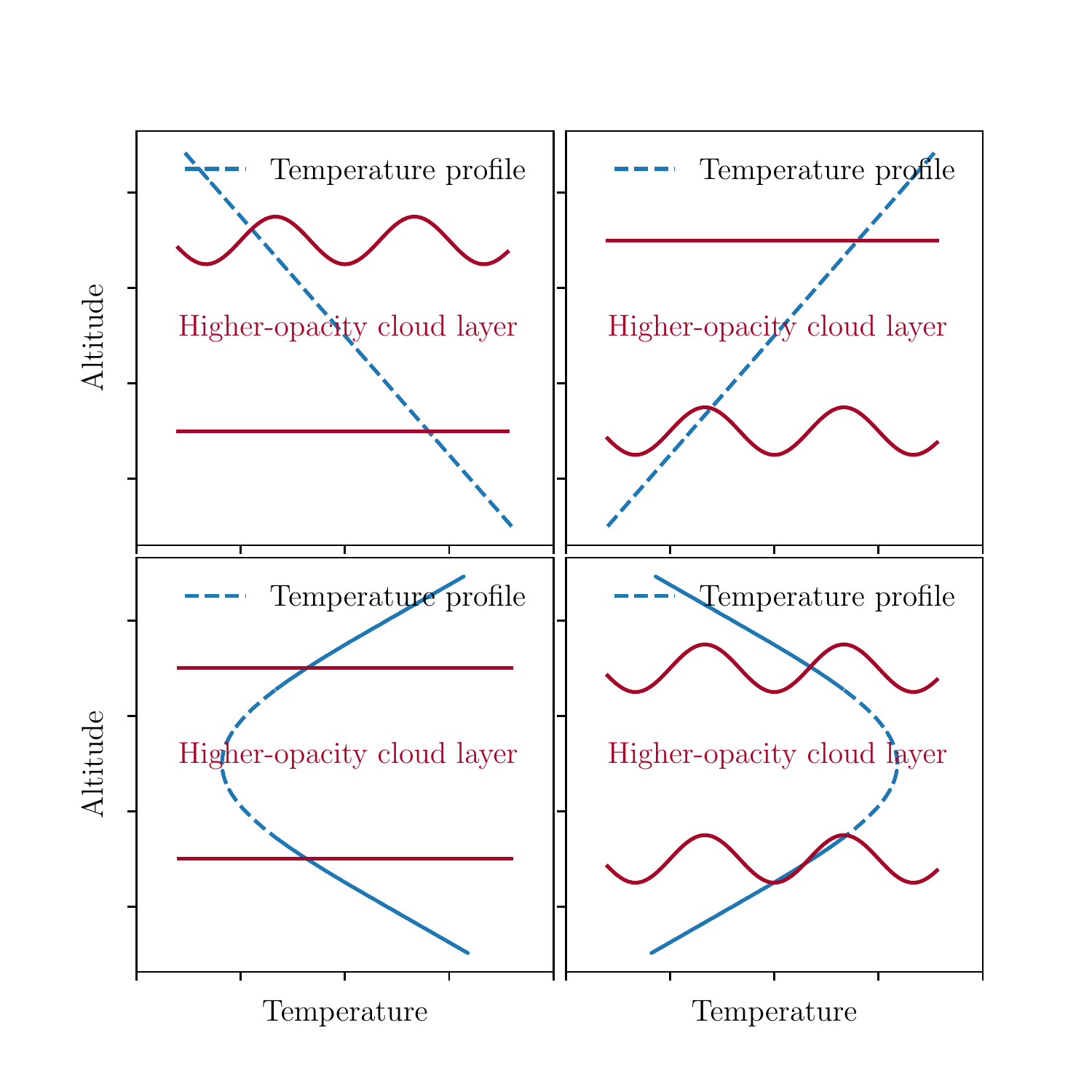}
\caption{Schematics of the possible stable/unstable situations for a higher-opacity
  cloud layer depending on the temperature structure of the
  atmosphere.} \label{fig:cloud_structure}
\end{centering}
\end{figure}

We schematically summerize in Fig.~\ref{fig:cloud_structure} all the different
possibilities of the stable/unstable regimes assuming the presence of a higher-opacity
cloud layer in an atmosphere with different temperature structures. Essentially
the top part of the cloud layer is unstable if the temperature decreases with
altitude, stable with a temperature inversion; the base is stable if the
temperature decreases with height and unstable with a temperature
inversion. This might lead to the cloud cover being patchy. Interestingly there
is only one case in which a cloud cover would be 
stable: negative temperature gradient at the base and a temperature inversion at
the top. We discuss in the next section the possible implications for cloud
covers in different objects.

\begin{table}
  \centering
  \caption{Parameters used to  estimate the radiative  and advective timescales
  plotted in Fig.~\ref{fig:timescales}.}\label{tab:param-timescales}
\begin{tabular}{l|l|l|l|l}
  & $c_p$ [cgs] & $T_\mathrm{1bar}$ [K]& u$_\mathrm{wind}$ [cgs] & $R_p$ [cgs] \\
  \hline
  \hline
  HD209458b & $1.5\times 10^{8}$& 1500 & $3.5\times 10^5$& $9.8 \times 10^9$\\
  2M 1047 & $1.5\times 10^{8}$& 900 & $6\times 10^4$& $5.5 \times 10^9$\\
  Venus & $1.16\times 10^{7}$& 360 & $1\times 10^4$& $6 \times 10^8$\\
  Earth & $1.1\times 10^{7}$& 290 & $1\times 10^4$& $6.3 \times 10^8$\\
  Jupiter & $1.3\times 10^{8}$& 150 & $1.5\times 10^4$& $7 \times 10^9$\\
  Saturn & $1.3\times 10^{8}$& 120 & $4\times 10^4$& $5.8 \times 10^9$\\
  Uranus & $8\times 10^{7}$& 75 & $2\times 10^4$& $2.5 \times 10^9$\\
  Neptune & $8\times 10^{7}$& 70 & $4\times 10^4$& $2.4 \times 10^9$
\end{tabular}
\end{table}

\begin{figure}[btp]
\begin{centering}
\includegraphics[width=1.0\linewidth]{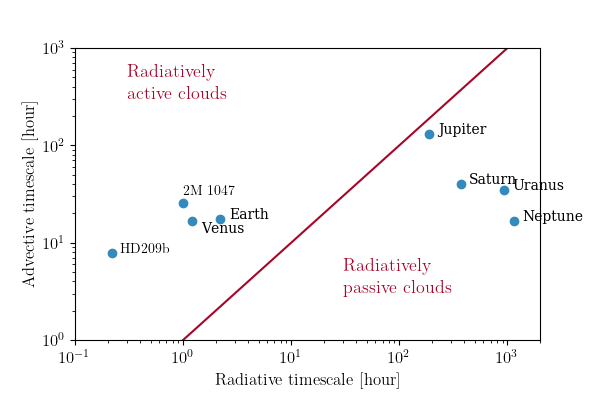}
\caption{Comparison between the radiative and advective timescale for HD209458b,
the T dwarf 2M 1047, Venus, Earth, Jupiter, Saturn, Uranus, and
Neptune.} \label{fig:timescales}
\end{centering}
\end{figure}

\begin{figure}[btp]
\begin{centering}
\includegraphics[width=1.0\linewidth]{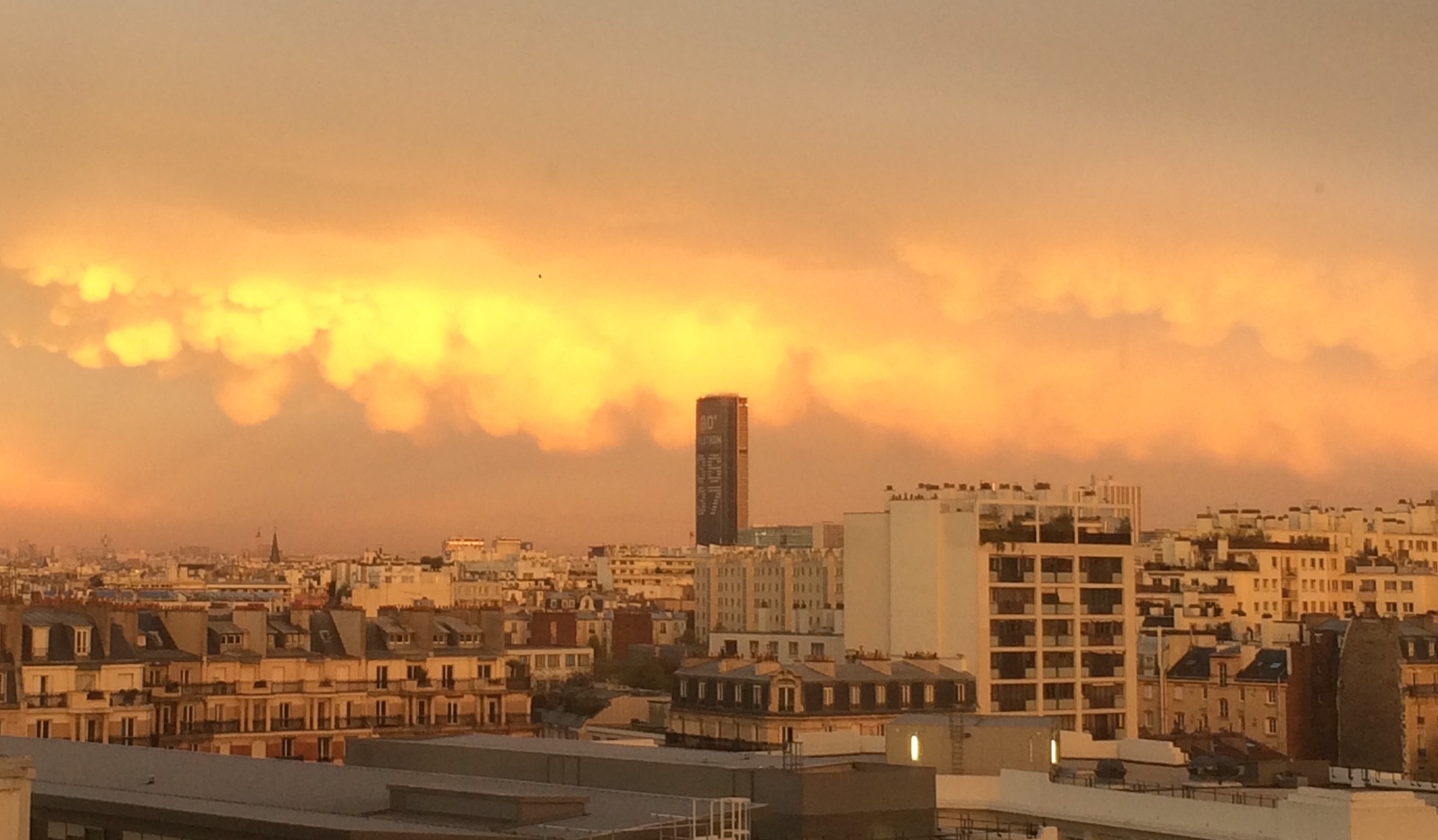}
\caption{Appearance of mammatus clouds over Montparnasse tower in Paris on the
  21st of November 2016.} \label{fig:mammatus}
\end{centering}
\end{figure}

\begin{figure}[btp]
\begin{centering}
\includegraphics[width=1.0\linewidth]{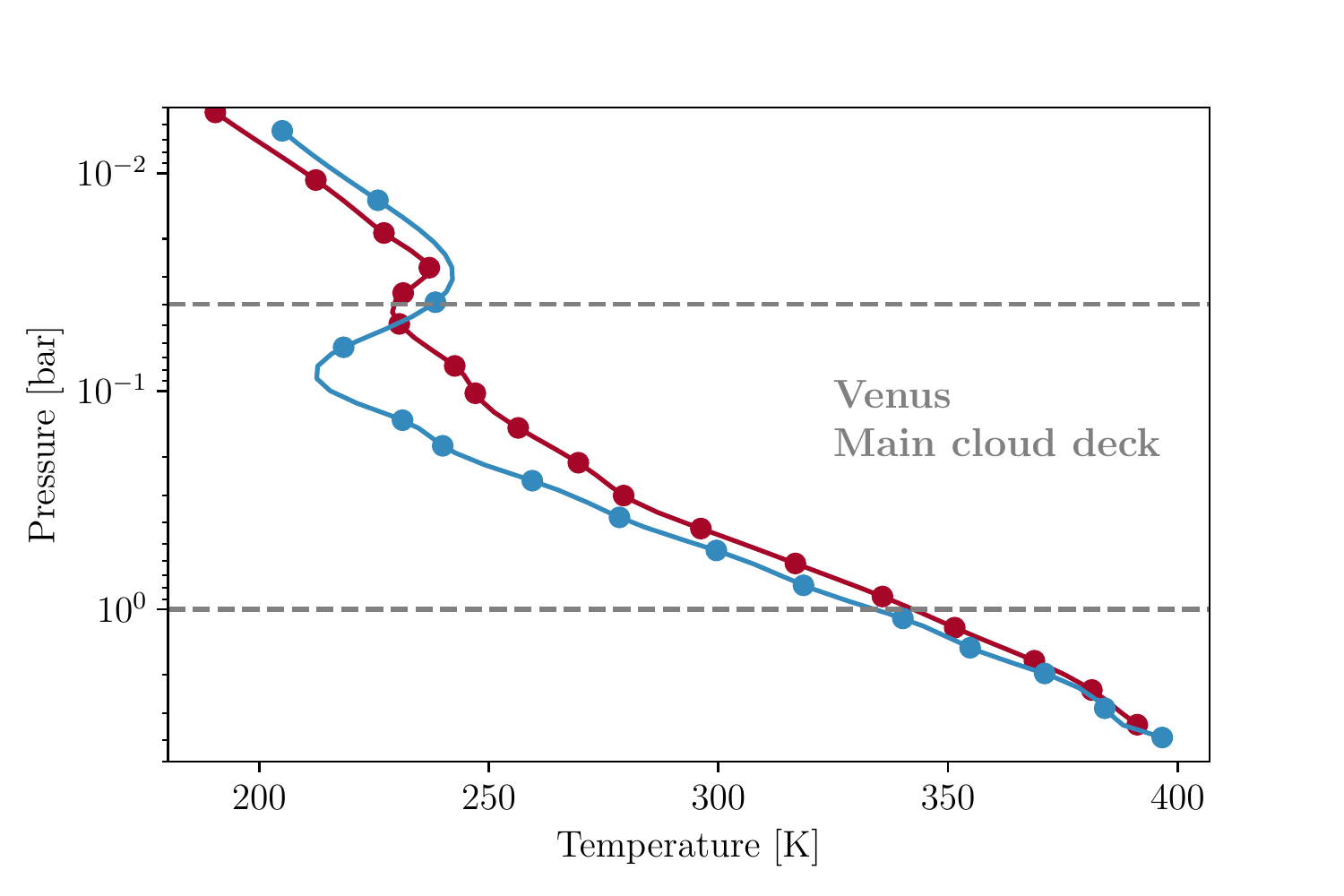}
\caption{Pressure/temperature profiles of the atmosphere of Venus measured by
  the Pioneer probes \citep{seiff:1983} and location of the main cloud deck
  \citep{formisano:2006}.} \label{fig:venus_profile}
\end{centering}
\end{figure}

\section{Interpretation on the structure of cloud covers}\label{sec:clouds}

\subsection{Cold ice- and gas-giant planets}

A first important distinction to make is to check if clouds are expected to be
radiatively active or passive, meaning that the radiative timescale is smaller
or larger than the advective timescale. If the radiative timescale is small,
the RRTI will develop quickly compared to other dynamical effects and will
strongly impact the shape of the clouds. if this timescale is long, wind and
convection will impact the higher-opacity layer before the radiative instability
develops and clouds can be seen as radiatively passive scalar following the
dynamics (ignoring for the moment condensation/evaporation see
Sect.~\ref{sec:conclusions}). We estimate in Fig.~\ref{fig:timescales} the
radiative timescale and advective timescale for the cloudy solar-system planets,
the extrasolar giant planet HD209458b and the T dwarf 2M 1047. We estimate the
radiative timescale  with
$\tau_\mathrm{rad}=c_p/\kappa\sigma T_\mathrm{1bar}^3$ assuming a cloud opacity
at $\kappa$= 
1~cm$^2$/g and a temperature from the pressure/temperature structure at 1
bar.
The global advective timescale is estimated with
$\tau_\mathrm{adv}=R_p/u_\mathrm{wind}$ with $R_p$ the radius of the object and
$u_\mathrm{wind}$ the typical wind velocities in the atmosphere. The wind
velocity for 2M 1047 is taken from the recent wind measurement in
\citet{Allers:2020}. All the values used to get these estimations are given in 
Tab.~\ref{tab:param-timescales}. We emphasize that this is an order-of-magnitude
estimation to assess roughly the  relative importance between radiative and
dynamical effects.

We can see that the cold ice- and gas-giant planets in the Solar System are
essentially in the regime in which the advective timescale is smaller than the
radiative timescale which implies that clouds are essentially radiatively
passive and follow the flow as a passive scalar. The radiative timescale is much
smaller for the giants than for Earth and Venus because they are colder, but
also because the specific heat capacity in these objects is around one order of
magnitude larger. As a result the radiative timescale is almost two orders of
magnitude larger. Surprisingly, HD209b and the T dwarf 2M 1047 fall in a similar
regime as Earth and Venus, mainly because the radiative timescale is small
because of the high temperature of these objects. Although the solar-system
giant planets are often seen as our closest analog to hot giant exoplanets and
brown dwarfs, they seem to be bad proxies to understand the cloud dynamics on
these objects compared to Earth and Venus, at least as far as the RRTI is
concerned. Similarly to the solar-system giants, we can expect that cold exoplanets
and cold Y dwarfs like Wise 0855 will also have relatively radiatively passive
clouds.

\subsection{Earth and Venus}

The dynamics of clouds in the atmosphere of Earth and Venus provides a possible
test bench for a mechanism such as the RRTI by comparing the correlation between
stability/instability and temperature gradients to Fig.~\ref{fig:cloud_structure}.

Mammatus clouds in Earth atmopshere are opaque lobes hanging at
the base of a cloud, typically beneath the anvil part of cumulonimbus when they
overshoot the tropopause or also at the base of thunderstorm clouds
\citep[see Fig.~\ref{fig:mammatus} and][]{winstead:2001}. The undestanding of
the formation mechanism of these
unusual cloud structures remains a challenge and many processes have been
proposed so far based on evaporative cooling, cloud-base radiative heating,
dynamical instabilities (Kelvin-Helmholtz, ``standard'' Rayleigh-Taylor) and many others
\citep{schultz:2006,schultz:2007,garrett:2010}. However none of these processes
seem to provide a definite answer since they do not seem to always lead to the
formation of mammated clouds. Yet among all the observed properties of the
mammatus, the most shared is certainly the correlation between their formation
and the presence of a temperature inversion. This correlation has been reported
as early as 1911 by Clayton using kite soundings at the Harvard College
Astronomical Observatory: ''In every case where the lower surface of a cloud is
mammated, the cloud is found above an inverted gradient of temperature''
\citep{clayton:1911}.

An other interesting case is the main cloud
cover in the atmosphere of Venus. Since a stable cloud deck is present and
prevent us from seeing the surface at visible wavelength, the RRTI would predict this
cloud cover to be stable only if a temperature inversion is present at the top
of the cloud deck. We provide in Fig.~\ref{fig:venus_profile} the
pressure/temperature
profiles measured by two of the four Pioneer probes \citep{seiff:1983} and the
location of the main cloud deck \citep{formisano:2006}. It does appear that a
temperature inversion is present at the location of the top boundary of the
clouds explaining the stability to the RRTI and the formation of the cloud cover on
the dayside of the planet. Yet such temperature inversions are not possible
without irradiation on the nightside of the planet. We would therefore expect
the top cloud cover to be destabilized there. This is also what has been
observed by Venus Express with relatively recent analysis of the cloud dynamics
on the nightside of the planet \citep{peralta:2017}. While the cloud cover is
smooth and follows super-rotation on the dayside (well reproduced by global
circulation models (GCM) e.g. \citet{lebonnois:2016,garate:2018}), a more
complex dynamics and cloud structure with
stationary wave patterns seems to be present on the nighside and is not
well-reproduced by GCM simulations. 

Interestingly, both types of phenomena seem to be qualitatively in agreement
with our predictions for the behavior of the RRTI as a function of the temperature
gradient (see Fig.~\ref{fig:cloud_structure}): the base of a cloud cover would
be unstable to the RRTI when a temperature inversion is present (similar to the case
of Mammatus) and the top of the cloud cover would be stable with a temperature
inversion (dayside of Venus) and unstable with temperature decreasing with
height (nightside of Venus). More detailed studies are needed to confirm this
link, e.g. with local convective simulations such as
\citet{lefevre:2017,lefevre:2018} in which the RRTI mechanism could be
identified and characterized.

\section{Discussion and Conclusions}\label{sec:conclusions}

\subsection{Speculations for exoplanets}

Based on the insight provided by the RRTI and its possible role in the dynamics of
Earth and Venus clouds, we can provide expectations for the cloud
cover of irradiated rocky and giant exoplanets. Similarly to the dayside of
Venus we could expect a stable cloud cover to be possible only when irradiation
create a temperature inversion at the top of the clouds (this inversion can
be caused by the absorption of stellar light by the clouds themselves). If the
cloud cover can grow to a large vertical extension on the dayside, it can then
be advected on the nightside and survive the instabilities created at the top by
the disappearance of the temperature inversion and consequently cover the entire
planet. This possibility will strongly depend on the rotation period and wind
velocity: if locally the clouds are submitted to a rapid day/night forcing they
could be rapidly destabilized before growing significantly at the planetary
scale, while for slow rotation/wind speed they could grow significantly on the
dayside and survive on the nightside. Our mechanism may contribute to shaping
certain cloud properties on Earth and Venus and in general on
irradiated exoplanets, although multiple mechanisms are
crucial in determining the weather system on Earth and Venus.

\begin{figure}[btp]
\begin{centering}
\includegraphics[width=1.0\linewidth]{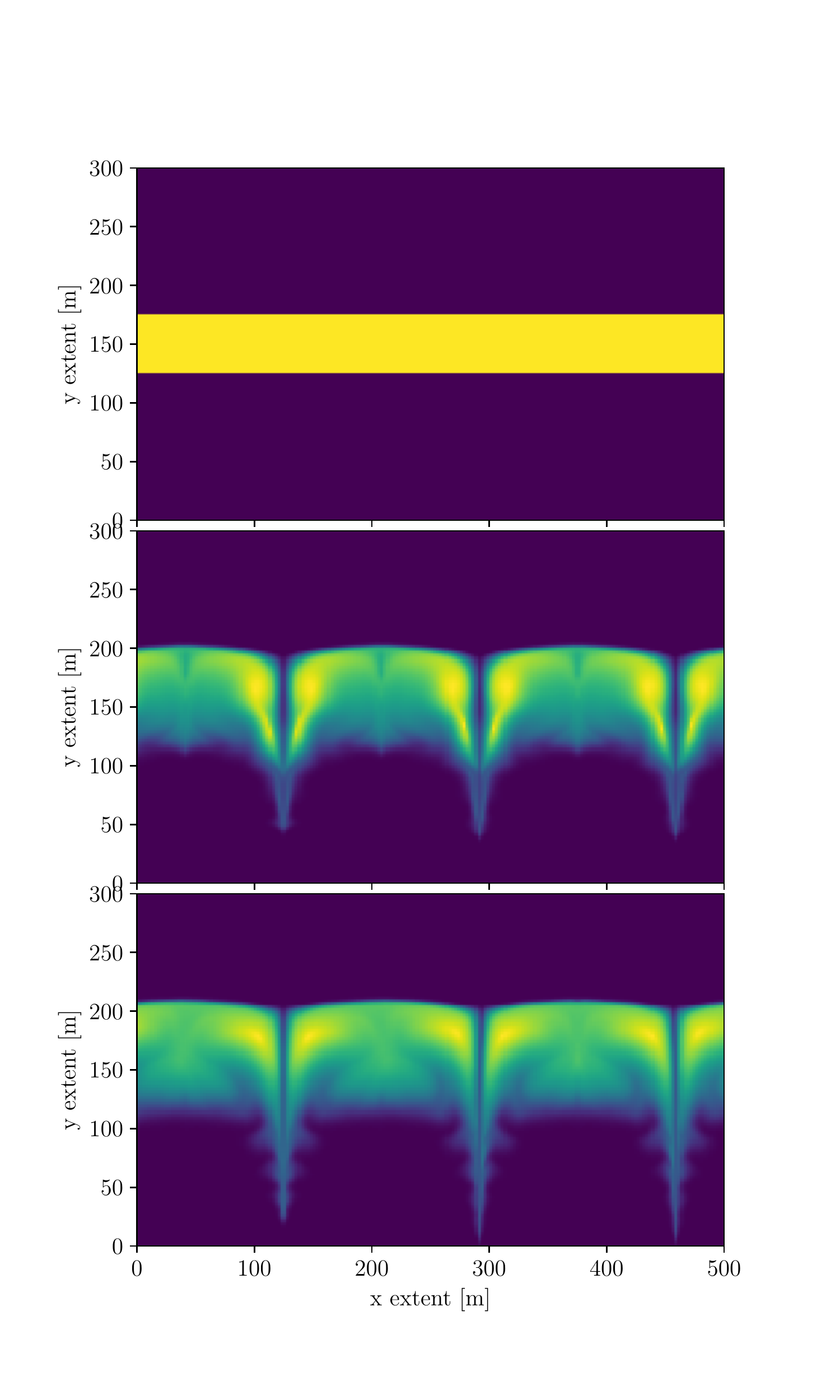}
\caption{Opacity tracer in a simulation with an unstable cloud cover. The top
  panel is the initial state, the middle at $t=150$s and, the bottom at $t=300$s.} \label{fig:sim_cover}
\end{centering}
\end{figure}

For isolated objects like brown dwarfs or low-irradiated hot exoplanets like the
one usually observed by direct imaging, the formation of temperature inversions
appears difficult and is not really expected to happen. We therefore expect the
top of a cloud cover to be unstable to the RRTI everywhere in the atmosphere of these
objects. As a consequence of the RRTI, patchy cloud covers may be ubiquitous in both L and T
dwarfs and not only at the L/T transition. To illustrate this point, we
  show in Fig. \ref{fig:sim_cover} a simulation with an unstable cloud layer
  leading to patches of opacity. We perform this simulation on longer time-scale
  to probe the saturated steady state, the solution does not evolve much between
  150 and 300 seconds which is therefore much longer than the turnover timescale
  associated to the initial transitional phase.
In that context, 1D atmospheric
models with homogeneous cloud cover may not be a sufficiently realistic
approximation. 

\subsection{Conclusions}

By using an analytical stability analysis and 2D radiative-hydrodynamical
simulations, we have shown the existence of an instability of an opacity
discontinuity between a lower-opacity and a higher-opacity medium: 
the Radiative Rayleigh-Taylor instability, a particular case of the general
diabatic Rayleigh-Taylor instability.

\begin{itemize}
\item In an atmosphere with a negative vertical gradient of temperature, a
  lower-opacity medium on top of a higher-opacity medium is unstable and a
  higher-opacity medium on top of a lower-opacity medium is stable.
\item In an atmosphere with a positive vertical gradient of temperature, the
  behavior is inverted: a lower-opacity medium on top of a higher-opacity medium is
  stable and an higher-opacity medium on top of a lower-opacity medium is unstable.
\item Applied to a higher-opacity cloud layer, this mechanism predicts that  the base
  of a cloud can be unstable in the presence of a temperature inversion and a
  cloud cover can only be stable if a temperature inversion is present at the top.
  \end{itemize}

This mechanism could shed some light on the interpretation of several cloud
structures in different objects:

\begin{itemize}
\item The radiative timescale of the ice- and gas-giant solar-system planets and
  cold exoplanets/Y dwarfs is long. In that context clouds can be expected to be
  radiatively passive and they essentially follow the dynamics as long as only
  radiative transfer is concerned.
\item The RRTI predicts that the base of a cloud is unstable with a
  temperature inversion, this could offer a possible mechanism to explain the
  formation of mammatus clouds in Earth and earth-like atmospheres. Furthermore
  a cloud cover can only be stable with a temperature inversion at the top which
  seems to be the case for Venus dayside temperature profiles. By generalizing
  this to irradiated exoplanets, we may expect stable large-scale cloud covers
  only in irradiated planets with slow rotation/wind speed, so that the cloud
  cover can grow on large scales on the dayside (and possibly be advected on the
  nightside).
\item Isolated and low-irradiated objects like brown dwarfs and hot exoplanets
  observed by direct imaging are not expected to have temperature inversions. In
  that context patchy cloud covers may be ubiquitous in their atmospheres.
  \end{itemize}

We have of course used a very simplified setup in this paper. Some limitations
regarding our current approach are:

\begin{itemize}
\item We partly neglected condensation/evaporation leading to the
  formation/destruction of the higher-opacity material and irradiation. We only
  explored with a simplified thermal forcing the possible formation of a
  temperature inversion because of evaporative cooling or irradiation. As a
    consequence, we have ignored the feedback of the instability on the forcing
    causing the temperature inversion.
\item We used a simplified approach for radiative transfer: a grey model
  neglecting scattering. We also used the two-stream
  approximation in column which neglects lateral radiative coupling.
  \item We used 2D simulations: further studies should look at this instability
    in 3D and assess its impact on a opaque layer of finite thickness.
    Longer timescales are also needed to probe the turbulent steady state.
\end{itemize}

Some of these limitations might not be important, e.g. condensation for the
formation of the higher-opacity material might happen in a first phase and be relatively
negligible in the following evolution governed by radiation
hydrodynamics. Scattering will probably not have much impact when the radiative
balance is mainly in the infrared, and in any case it does not impact radiative
heating and cooling. However a better modeling of evaporation, irradiation, and
3D radiative transfer \citep[e.g. with the M1 model][]{gonzalez:2007} are
certainly the next steps toward a better understanding of the RRTI and cloud dynamics
in the context of exoplanets and brown dwarfs. With the upcoming arrival of the
{\it James Webb Space Telescope} and observational data with a large spectral
coverage, these types of models might be essential for the understanding of
clouds in the atmosphere of giant and rocky exoplanets.

\begin{acknowledgements}
The authors are thankful to the anonymous referee for his/her valuable
  comments that help improving this manuscript. PT acknowledges supports by the
European Research Council under Grant 
Agreement ATMO 757858. PT also thanks A. Barazzutti for providing
Fig.~\ref{fig:mammatus}, for her natural curiosity about natural phenomena
without which this work would not have been possible, and her inconditional
support over the years.
\end{acknowledgements}

\bibliographystyle{aa}
\bibliography{main}

\begin{thebibliography}{29}
\expandafter\ifx\csname natexlab\endcsname\relax\def\natexlab#1{#1}\fi

\bibitem[{{Allers} {et~al.}(2020){Allers}, {Vos}, {Biller}, \&
  {Williams}}]{Allers:2020}
{Allers}, K.~N., {Vos}, J.~M., {Biller}, B.~A., \& {Williams}, P. K.~G. 2020,
  Science, 368, 169

\bibitem[{{Chandrasekhar}(1961)}]{chandrasekhar:1961}
{Chandrasekhar}, S. 1961, {Hydrodynamic and hydromagnetic stability}

\bibitem[{Clayton(1911)}]{clayton:1911}
Clayton, H.~H. 1911, Annal. Astron. Observatory Harvard, 68, 170

\bibitem[{Formisano {et~al.}(2006)Formisano, Al., \&
  Lebonnois}]{formisano:2006}
Formisano, V., Al., E., \& Lebonnois, S. 2006, {Planetary and Space Science},
  54, 1298

\bibitem[{{Fortney} {et~al.}(2008){Fortney}, {Lodders}, {Marley}, \&
  {Freedman}}]{fortney:2008}
{Fortney}, J.~J., {Lodders}, K., {Marley}, M.~S., \& {Freedman}, R.~S. 2008,
  \apj, 678, 1419

\bibitem[{{Garate-Lopez} \& {Lebonnois}(2018)}]{garate:2018}
{Garate-Lopez}, I. \& {Lebonnois}, S. 2018, \icarus, 314, 1

\bibitem[{Garrett {et~al.}(2010)Garrett, Schmidt, Kihlgren, \&
  Cornet}]{garrett:2010}
Garrett, T.~J., Schmidt, C.~T., Kihlgren, S., \& Cornet, C. 2010, Journal of
  the Atmospheric Sciences, 67, 3891

\bibitem[{{Gierasch} {et~al.}(1973){Gierasch}, {Ingersoll}, \& {Terry
  Williams}}]{gierasch:1973}
{Gierasch}, P.~J., {Ingersoll}, A.~P., \& {Terry Williams}, R. 1973, \icarus,
  19, 473

\bibitem[{{Gonz{\'a}lez} {et~al.}(2007){Gonz{\'a}lez}, {Audit}, \&
  {Huynh}}]{gonzalez:2007}
{Gonz{\'a}lez}, M., {Audit}, E., \& {Huynh}, P. 2007, \aap, 464, 429

\bibitem[{{Helling} {et~al.}(2019){Helling}, {Iro}, {Corrales}, {Samra},
  {Ohno}, {Alam}, {Steinrueck}, {Lew}, {Molaverdikhani}, {MacDonald},
  {Herbort}, {Woitke}, \& {Parmentier}}]{helling:2019}
{Helling}, C., {Iro}, N., {Corrales}, L., {et~al.} 2019, arXiv e-prints,
  arXiv:1906.08127

\bibitem[{Kokhanovsky(2004)}]{kokhanovsky:2004}
Kokhanovsky, A. 2004, Earth-Science Reviews, 64, 189

\bibitem[{{Kreidberg} {et~al.}(2014){Kreidberg}, {Bean}, {D{\'e}sert},
  {Benneke}, {Deming}, {Stevenson}, {Seager}, {Berta-Thompson}, {Seifahrt}, \&
  {Homeier}}]{kreidberg:2014}
{Kreidberg}, L., {Bean}, J.~L., {D{\'e}sert}, J.-M., {et~al.} 2014, \nat, 505,
  69

\bibitem[{{Lebonnois} {et~al.}(2016){Lebonnois}, {Sugimoto}, \&
  {Gilli}}]{lebonnois:2016}
{Lebonnois}, S., {Sugimoto}, N., \& {Gilli}, G. 2016, \icarus, 278, 38

\bibitem[{{Lef{\`e}vre} {et~al.}(2018){Lef{\`e}vre}, {Lebonnois}, \&
  {Spiga}}]{lefevre:2018}
{Lef{\`e}vre}, M., {Lebonnois}, S., \& {Spiga}, A. 2018, Journal of Geophysical
  Research (Planets), 123, 2773

\bibitem[{{Lef{\`e}vre} {et~al.}(2017){Lef{\`e}vre}, {Spiga}, \&
  {Lebonnois}}]{lefevre:2017}
{Lef{\`e}vre}, M., {Spiga}, A., \& {Lebonnois}, S. 2017, Journal of Geophysical
  Research (Planets), 122, 134

\bibitem[{{Lines} {et~al.}(2018){Lines}, {Mayne}, {Boutle}, {Manners}, {Lee},
  {Helling}, {Drummond}, {Amundsen}, {Goyal}, {Acreman}, {Tremblin}, \&
  {Kerslake}}]{lines:2018}
{Lines}, S., {Mayne}, N.~J., {Boutle}, I.~A., {et~al.} 2018, \aap, 615, A97

\bibitem[{{Mihalas} \& {Mihalas}(1984)}]{mihalas:1984}
{Mihalas}, D. \& {Mihalas}, B.~W. 1984, {Foundations of radiation
  hydrodynamics}

\bibitem[{{Padioleau} {et~al.}(2019){Padioleau}, {Tremblin}, {Audit},
  {Kestener}, \& {Kokh}}]{padioleau:2019}
{Padioleau}, T., {Tremblin}, P., {Audit}, E., {Kestener}, P., \& {Kokh}, S.
  2019, \apj, 875, 128

\bibitem[{{Parmentier} {et~al.}(2016){Parmentier}, {Fortney}, {Showman},
  {Morley}, \& {Marley}}]{parmentier:2016}
{Parmentier}, V., {Fortney}, J.~J., {Showman}, A.~P., {Morley}, C., \&
  {Marley}, M.~S. 2016, \apj, 828, 22

\bibitem[{{Peralta} {et~al.}(2017){Peralta}, {Hueso}, {S{\'a}nchez-Lavega},
  {Lee}, {Mu{\~n}oz}, {Kouyama}, {Sagawa}, {Sato}, {Piccioni}, {Tellmann},
  {Imamura}, \& {Satoh}}]{peralta:2017}
{Peralta}, J., {Hueso}, R., {S{\'a}nchez-Lavega}, A., {et~al.} 2017, Nature
  Astronomy, 1, 0187

\bibitem[{Schultz {et~al.}(2007)Schultz, Kanak, Straka, Trapp, Gordon, Zrnic,
  Bryan, Durant, Garrett, Klein, \& Lilly}]{schultz:2007}
Schultz, D., Kanak, K., Straka, J., {et~al.} 2007, Bulletin of the American
  Meteorological Society, 88, 146

\bibitem[{Schultz {et~al.}(2006)Schultz, Kanak, Straka, Trapp, Gordon,
  Zrni{\'c}, Bryan, Durant, Garrett, Klein, \& et~al.}]{schultz:2006}
Schultz, D.~M., Kanak, K.~M., Straka, J.~M., {et~al.} 2006, Journal of the
  Atmospheric Sciences, 63, 2409

\bibitem[{{Seiff}(1983)}]{seiff:1983}
{Seiff}, A. 1983, {Thermal structure of the atmosphere of Venus.}, 215--279

\bibitem[{{Sing} {et~al.}(2016){Sing}, {Fortney}, {Nikolov}, {Wakeford},
  {Kataria}, {Evans}, {Aigrain}, {Ballester}, {Burrows}, {Deming},
  {D{\'e}sert}, {Gibson}, {Henry}, {Huitson}, {Knutson}, {Lecavelier Des
  Etangs}, {Pont}, {Showman}, {Vidal-Madjar}, {Williamson}, \&
  {Wilson}}]{sing:2016}
{Sing}, D.~K., {Fortney}, J.~J., {Nikolov}, N., {et~al.} 2016, \nat, 529, 59

\bibitem[{{Tan} \& {Showman}(2019)}]{tan:2019}
{Tan}, X. \& {Showman}, A.~P. 2019, \apj, 874, 111

\bibitem[{{Tan} \& {Showman}(2021{\natexlab{a}})}]{tan:2021a}
{Tan}, X. \& {Showman}, A.~P. 2021{\natexlab{a}}, \mnras, 502, 678

\bibitem[{{Tan} \& {Showman}(2021{\natexlab{b}})}]{tan:2021b}
{Tan}, X. \& {Showman}, A.~P. 2021{\natexlab{b}}, \mnras, 502, 2198

\bibitem[{{Tremblin} {et~al.}(2019){Tremblin}, {Padioleau}, {Phillips},
  {Chabrier}, {Baraffe}, {Fromang}, {Audit}, {Bloch}, {Burgasser}, {Drummond},
  {Gonz{\'a}lez}, {Kestener}, {Kokh}, {Lagage}, \& {Stauffert}}]{tremblin:2019}
{Tremblin}, P., {Padioleau}, T., {Phillips}, M.~W., {et~al.} 2019, \apj, 876,
  144

\bibitem[{Winstead {et~al.}(2001)Winstead, Verlinde, Arthur, Jaskiewicz,
  Jensen, Miles, \& Nicosia}]{winstead:2001}
Winstead, N.~S., Verlinde, J., Arthur, S.~T., {et~al.} 2001, Monthly Weather
  Review, 129, 159

\end{thebibliography}

\begin{appendix}

\section{Linear stability analysis of of the diabatic Rayleigh-Taylor instability}\label{app:rrti}

\subsection{General equations}

For the linear stability analysis, we will use the Boussinesq approximation
following  \citet{tremblin:2019}. We
 decompose all the fields as e.g.
$\rho(t,x,y,z) = \rho_0(z)+\delta\rho(t,x,y,z)$ and the background state is given by the
static equations:

\begin{eqnarray}
  \frac{\partial P_0}{\partial z} &=& -\rho_0 g \cr
  R(X_0,T_0) &=& 0 \cr
  \vec{u} &=& 0 \cr
  H(X_0,T_0) &=& 0
\end{eqnarray}

with $R(X,T)$ the source term in the advection/reaction equation of the fluid
concentration $X$, and $\rho C_p H(X,T)$ the source term in the energy equation i.e.
thermal conduction or radiative transfer heating rate.

The linearized Boussinesq approximation leads to the following system:

\begin{eqnarray}
\vec{\nabla}\left(\vec{\delta u}\right) &=& 0 \cr
\frac{\partial \delta X}{\partial t} +
\vec{\delta u}\cdot\vec{\nabla}\left(X_0\right) &=&  \frac{\partial R}{\partial X}
\delta X+ \frac{\partial R}{\partial T} \delta T \cr
\frac{\partial\rho_0 \vec{\delta u}}{\partial t} + \vec{\nabla}\left(\delta P\right)
 - \delta\rho \vec{g} &=& 0\cr
\frac{\partial \delta T}{\partial t} + T_0 \vec{\delta
  u}\cdot \vec{\nabla}\log \theta_0
 &=&  \frac{\partial H}{\partial X} \delta X
 + \frac{\partial H}{\partial T} \delta T
\end{eqnarray}
with $\theta_0$ the potential temperature
$T_0(P_\mathrm{ref}/P_0)^{(\gamma-1)/\gamma}$ ($P_\mathrm{ref}$ is a reference 
pressure and $\gamma$ the adiabatic index). The system is closed with the
equation of state for an ideal gas (by removing $\delta P$ in 
the Boussinesq limit) :
\begin{equation}
0 = \frac{\delta \rho}{\rho_0} + \frac{\delta T}{T_0} -
\frac{\partial \log \mu_0}{\partial X} \delta X
\end{equation}
and we consider in the rest of the analysis an interface between two sub-domains
located in $z=0$. Before going to the Boussinesq regime, we recall first the
demonstration in the 
incompressible regime, without composition nor source terms in order to
highlight the modification of the demonstration for the Boussinesq limit.

\subsection{Incompressible limit}

In the incompressible case, the system reduces to:
\begin{eqnarray}
\vec{\nabla}\left(\vec{\delta u}\right) = 0 \cr
\frac{\partial\rho_0 \vec{\delta u}}{\partial t} + \vec{\nabla}\left(\delta P\right)
 - \delta\rho \vec{g} = 0\cr
\end{eqnarray}
And we can introduce an extra equation on the advection of density:
\begin{equation}
\frac{\partial \delta \rho}{\partial t} + \vec{\delta u} \vec{\nabla}\rho_0 = 0
\end{equation}
Then we assume the form $\exp(\omega t +i(k_x x + k_y y))$ for the disturbance:
\begin{eqnarray}
 i k_x \delta u + i k_y \delta v + \partial_z \delta w = 0 \cr
\omega \rho_0 \delta u + i k_x \delta P = 0\cr
\omega \rho_0 \delta v + i k_y \delta P = 0\cr
\omega \rho_0 \delta w +  \partial_z \delta P + \delta \rho g = 0\cr
\omega \delta \rho + \delta w \partial_z \rho_0 = 0
\end{eqnarray}
which reduces to (defining $k^2 = k_x^2+k_y^2)$:
\begin{eqnarray}
 k^2 \delta P + \omega \rho_0 \partial_z \delta w = 0 \cr
\omega \rho_0 \delta w +  \partial_z \delta P + \delta \rho g = 0\cr
\omega \delta \rho + \delta w \partial_z \rho_0 = 0
\end{eqnarray}
We then need to find the differential equation for $\delta w$:

\begin{equation}\label{eq:dif}
\omega \rho_0 \delta w  - \omega \partial_z \left( \rho_0 \partial_z\delta w
\right)/k^2 + \delta \rho g = 0
\end{equation}
with the relation that links $\delta \rho$ and $\delta w$ in the incompressible
limit:
\begin{equation}
\omega \delta \rho + \delta w \partial_z \rho_0 = 0
\end{equation}
We explicitely keep the relation between $\delta \rho$ and $\delta w$ here,
because this is what is going to change in the Boussinesq limit with source terms.
If we assume that in each subdomain, the density is constant, the differential
equation is then

\begin{equation}
  \partial_z^2\delta w - k^2 \delta w = 0,
\end{equation}

  which has the
solution $\delta w(0^\pm) e^{\mp k z}$ and assuming the continuity of velocity
at the interface leads to $\delta w(0^\pm)=\delta w (0)$.

We now replace $\delta \rho$ in the differential equation Eq.~\ref{eq:dif} and
integrate around the interface between $z=\pm \epsilon$ to get the jump relation:

\begin{equation}
\omega^2 k^2\int_{-\epsilon}^{+\epsilon} \rho_0 \delta w dz-\omega^2[\rho_0\partial_z
  \delta w]^{+\epsilon}_{-\epsilon}-gk^2 \int_{-\epsilon}^{+\epsilon}\delta w
\partial_z \rho_0 dz = 0
\end{equation}
The first term can be decomposed as follow:
\begin{eqnarray}
\int_{-\epsilon}^{+\epsilon} \rho_0 \delta w dz &=&  \int_{-\epsilon}^{0} \rho_0
\delta w dz + \int_{0}^{+\epsilon} \rho_0 \delta w dz \cr
&\approx& -\epsilon \rho_- \delta w(0) + \epsilon \rho_+ \delta w(0)
\end{eqnarray}
which tends to zero as $\epsilon$ tends to zero. This result holds for any
term that cannot be written as a derivative of a discontinuous function: only
those terms (i.e. dirac functions) contribute to the jump relation.
The second term can be written as follow:
\begin{eqnarray}
[\rho_0\partial_z
  \delta w]^{+\epsilon}_{-\epsilon} &=& -k \rho_0(\epsilon) \delta w(\epsilon)
e^{-k\epsilon} - k\rho_0(-\epsilon) \delta w(-\epsilon)  e^{+k(-\epsilon)} \cr
&\rightarrow & -k\delta w(0)(\rho_-+\rho_+)
\end{eqnarray}
$\delta w$ can be removed from the integral in the third term since it is
continuous at the interface and we can perfom the integration similarly to the
second term to obtain the classical result: 

\begin{equation}
\omega^2 = gk \frac{\rho_+-\rho_-}{\rho_++\rho_-}
\end{equation}

We now introduce the  Boussinesq limit, without composition nor source terms.

\subsection{Boussinesq limit}\label{sect:bous}

In the Boussinesq limit, the differential equation in \ref{eq:dif} is the same
but the link between $\delta \rho$ and $\delta w$ is different. Assuming no
composition nor source terms, the Boussinesq limit add the following relations:

\begin{eqnarray}
  \omega \delta T + \delta w T_0\partial_z (\log \theta_0) &=& 0\cr
  \delta \rho/\rho_0  &=& -\delta T /T_0
\end{eqnarray}
which gives the following relation between $\delta \rho$ and $\delta w$:
\begin{equation}\label{eq:rho_bous}
\omega \delta \rho/\rho_0 = \delta w \partial_z (\log \theta_0)
\end{equation}
Note that the relation between
$\delta \rho$ and $\delta w$ is proportional to the instability criterion in the
continuous case ($\partial_z \log \theta_0 <0$). We assume the log-density
gradient and the log-potential 
temperature gradient constant in each subdomain. This assumption might not be
true depending on the equation of state but we will assume that the wavelength
of the perturbation is sufficiently small so that both gradients can be
considered constant. The differential equation Eq.~\ref{eq:dif} with
Eq.~\ref{eq:rho_bous} then become: 
\begin{eqnarray}\label{eq:dif_bous}
  \omega^2 \partial_z (\rho_0\partial_z \delta w)-\omega^2 k^2 \rho_0\delta w -g k^2 \rho_0
  \partial_z(\log \theta_0) \delta w &=& 0\cr
\partial_z^ 2\delta w +\partial_z(\log \rho_0) \partial_z \delta w -k^2(1 +
g\partial_z (\log \theta_0)/\omega^2)\delta w &=& 0
\end{eqnarray}
whose solution is $\delta w(0^\pm) e^{q^\pm z}$ with
\begin{equation}\label{eq:q}
  \begin{split}
  q^\pm &= -(\partial_z(log \rho_0))^\pm/2\\
  &\mp \sqrt{(\partial_z(\log \rho_0))^{\pm,2}/4 +k^2\left(1+g(\partial_z(\log
    \theta_0))^\pm/\omega^2\right)}
  \end{split}
\end{equation}
Note that $q^\pm$ is independant of $\omega$ if the subdomains are neutral to
buoyancy ($\partial_z \log \theta_0 = 0$).

To write the jump conditions we need
to write the non-conservative product $\rho_0 \partial_z \log \theta_0$ with
discontinuous variables only in the derivative (using the equation of
hydrostatic equilibrium)
\begin{equation}
\rho_0 \partial_z \log \theta_0 = -\partial_z \rho_0 - \rho_0^2 g/(P_0 \gamma)
\end{equation}
The second term can be discontinuous but involves no derivatives, therefore it
does not contribute to jump relations (by integration of this term on each side
of the inteface). Only the first term (equivalent to a dirac function)
contributes. The jump condition by
integration of Eq.~\ref{eq:dif_bous} between $\pm \epsilon $ is then
given by
\begin{equation}
\omega^2 [ \rho_0 \partial_z \delta w]^{+\epsilon}_{-\epsilon}+
      [\rho_0]^{+\epsilon}_{-\epsilon}\delta w(0)g k^2 = 0
\end{equation}
Hence the growth rate in the Boussinesq limit is given by:
\begin{equation}
\omega^2 = gk^2 \frac{\rho_+-\rho_-}{\rho_{-} q^- -\rho_+ q^+}
\end{equation}
Note that we recover the incompressible result if the log-density and
log-potential temperature gradients are negligible in Eq.~\ref{eq:q},
i.e. $q^\pm = \mp k$.

\subsection{Diabatic Rayleigh-Taylor instability}

Now we go back to the full system with composition and  source terms. The
equations we need to add now to Eq.~\ref{eq:dif} are

\begin{eqnarray}
  \omega \delta X + \delta w \partial_z X_0 &=& R_X \delta X + R_T \delta T\cr
  \omega \delta T + \delta w T_0 \partial_z(\log \theta_0) &=& H_T \delta T +
  H_X\delta X \cr
   \delta \rho/\rho_0 &=& - \delta T/T_0 + \partial_X(\log \mu_0) \delta X
\end{eqnarray}
In that case
\begin{eqnarray}
 (\omega^\prime_T -\omega)\delta w \partial_z X_0 &=& (H_T-\omega
  +H_X/(T_0 \partial_X \log \mu_0))\times\cr
  &&((R_X -\omega)\delta X + R_T \delta T)\cr
  (\omega^\prime_X -\omega)\delta w \partial_z(\log \theta_0) &=&
  ((R_X-\omega)/T_0+R_T\partial_X(\log \mu_0))\times\cr
  && ((H_T-\omega) \delta T +H_X\delta X) \cr
   \delta \rho/\rho_0 &=& - \delta T/T_0 + \partial_X(\log \mu_0) \delta X
\end{eqnarray}
with $\omega^\prime_X = R_X+R_T T_0 \partial_X \log\mu_0$ and $\omega^\prime_T =
H_T + H_X/(T_0 \partial_X \log\mu_0)$. This implies the relation between $\delta
\rho$ and $\delta w$:
\begin{eqnarray}
  \delta \rho/\rho_0  &=& \delta w (\alpha_T  \partial_z \log \mu_0
  - \alpha_X \partial_z(\log \theta_0))\cr
  \alpha_T &=& (\omega^\prime_T-\omega)/((H_T-\omega)(R_X-\omega)-H_X R_T) \cr
  \alpha_X &=& (\omega^\prime_X-\omega) /((H_T-\omega)(R_X-\omega)-H_X R_T)
\end{eqnarray}
We then assume $\alpha_T$ and $\alpha_X$ constant in the whole domain. We can then
redefine
\begin{eqnarray}\label{eq:rho_dia}
  \log \psi &=& -\alpha_X \log \theta_0 + \alpha_T \log \mu_0 \cr
   \delta \rho/\rho_0  &=& \delta w \partial_z(\log \psi)
\end{eqnarray}
The differential equation Eq.~\ref{eq:dif} with
Eq.~\ref{eq:rho_dia} then become:
\begin{eqnarray}\label{eq:dif_dia}
  \omega^2 \partial_z (\rho_0\partial_z \delta w)-\omega^2 k^2 \rho_0\delta w -g k^2 \rho_0
  \omega \partial_z(\log \psi) \delta w &=& 0\cr
\partial_z^ 2\delta w +\partial_z(\log \rho_0) \partial_z \delta w -k^2(1 +
g\partial_z (\log \psi)/\omega)\delta w &=& 0
\end{eqnarray}

In each subdomain we can solve the differential equation in $\delta w$ to get
$\delta w(0^\pm) e^{q^\pm z}$ with
\begin{eqnarray}\label{eq:q_dia}
  q^\pm &=& -(\partial_z(log \rho_0))^\pm/2\cr
  &\mp& \sqrt{(\partial_z(\log
  \rho_0))^{\pm,2}/4 +k^2\left(1+g (\partial_z(\log \psi))^\pm/\omega\right)}
\end{eqnarray}

Remarkably, we can get a discontinuity (and instability) in $\log \psi$ with a
continuous density at the interface. 
We will  therefore assume $\rho_0$
continuous at the interface. We also need to do this hypothesis in order to have a well-defined
jump relation for the term $\rho_0 \partial_z(\log\psi)$, with a discontinuous
density this is a non-conservative product that is a-priori not well defined. We
highlight that this hypothesis means that the interface is neutral to the
classic Rayleigh-Taylor instability but does not prevent the medium to be
stratified on each side of the interface (with non-zero
$\partial_z(\log\theta_0)^\pm$ in Eq.~\ref{eq:rho_dia} and \ref{eq:q_dia}).

The jump condition, then become by integration of
Eq.~\ref{eq:dif_dia} between $\pm \epsilon$:
\begin{equation}
\omega \rho_0  [ \partial_z \delta w ]^{+\epsilon}_{-\epsilon}-[\log
  \psi]^{+\epsilon}_{-\epsilon} \rho_0 \delta w(0)g k^2 = 0
\end{equation}
which gives the diabatic Rayleigh-Taylor growth rate for a continuous density:
\begin{equation}\label{eq:gr_dia}
\omega  = g k^2 \frac{\log \psi^+ - \log \psi^-}{q^+ - q^-}
\end{equation}

Note that this growth rate takes into account the stabilizing effect of
stratification in the medium $\partial_z(\log\theta_0)^\pm \neq 0$. In the limit
of very large stratification $q^\pm \rightarrow \mp k \sqrt{g (\partial_z(\log
  \psi))^\pm/\omega}$, which gives the following limit of the growth rate

\begin{equation}
\omega^{3/2}  = \sqrt{g} k \frac{\log \psi^+ - \log \psi^-}{\sqrt{(\partial_z(\log
  \psi))^+} +\sqrt{(\partial_z(\log \psi))^-} }
\end{equation}

In the limit of small source  terms $H_{X,T},R_{X,T} \rightarrow 0$, we have $(\partial_z(\log \psi))^\pm
\rightarrow( (\partial_z(\log \theta_0))^\pm - ( \partial_z(\log \mu_0))^\pm
)/\omega$, which shows that $\omega \rightarrow 0$ when $ \partial_z(\log
\theta_0))^\pm\rightarrow +\infty$.

\subsection{Radiative Rayleigh-Taylor instability}

Assuming $R = 0$ and no mean molecular weight gradient, we get:
\begin{eqnarray}
  \omega \delta X + \delta w \partial_z X_0 &=&0\cr
  \omega \delta T + \delta w T_0 \partial_z(\log \theta_0) &=& H_T \delta T +
  H_X\delta X \cr
   \delta \rho/\rho_0 &=& - \delta T/T_0 
\end{eqnarray}
assuming only a discontinuous composition $X$, we get
\begin{eqnarray}
  \log \psi &=& -H_X  X_0 /(\omega T_0 (H_T-\omega))\cr
   \delta \rho/\rho_0  &=&  \delta w \partial_z(\log \psi)
\end{eqnarray}
which gives the growth rate,  assuming continuous $\rho_0$:
\begin{equation}
\omega^2  = g k^2 \frac{H_X}{T_0 (H_T-\omega)} \frac{X_0^- -X_0^+}{q^+ - q^-} 
\end{equation}
We can simplify this expression assuming $\omega\ll H_T$ and neglecting the
log-density and $\log 
\psi$ gradients such that $q^\pm=\mp k$. In this limit, the radiative
Rayleigh-Taylor growth rate is simply given by:
\begin{equation}
\omega^2  = \frac{g k}{2} \frac{H_X}{T_0 H_T}\left( X_0^- - X_0^+\right)
\end{equation}
A discontinuity of composition associated to a source term that depends on
composition (i.e. a discontinuity in $X H_X$) can 
therefore lead to an instability similar to the adiabatic Rayleigh-Taylor
instability even if the density is continuous at  the interface.

\end{appendix}

\end{document}